\documentclass[fleqn,usenatbib]{mnras}

\usepackage{newtxtext,newtxmath}
\usepackage{xcolor}
\usepackage{soul}
\usepackage{ulem}

\usepackage[T1]{fontenc}

\DeclareRobustCommand{\VAN}[3]{#2}
\let\VANthebibliography\thebibliography
\def\thebibliography{\DeclareRobustCommand{\VAN}[3]{##3}\VANthebibliography}

\usepackage{graphicx}	
\usepackage{amsmath}	

\usepackage{xcolor}

\title[Dwarf Galaxy Quenching in the EoR]{MEGATRON: Dwarf Galaxy Quenching in the Epoch of Reionization}

\author[D. Attard et al.]{
David Attard$^{1,2}$,
Ilian T. Iliev$^{1}$,
Martin P. Rey$^{3}$,
Harley Katz$^{4}$,
Corentin Cadiou$^{5}$ and
Nicholas Choustikov$^{6}$
\\
$^{1}$Department of Physics \& Astronomy, University of Sussex, Brighton, BN1 9QH, UK\\
$^{2}$DISCnet Centre for Doctoral Training, University of Sussex, Brighton, BN1 9QH, United Kingdom\\
$^{3}$Department of Physics, University of Bath, Claverton Down, Bath BA2 7AY, UK\\
$^{4}$Kavli Institute for Cosmological Physics, University of Chicago, Chicago IL 60637, USA\\
$^{5}$Institut d’Astrophysique de Paris, Sorbonne Universit\'e, CNRS, UMR 7095, 98 bis bd Arago, 75014 Paris, France\\
$^{6}$Sub-department of Astrophysics, University of Oxford, Keble Road, Oxford OX1 3RH, United Kingdom\\
}

\date{Accepted XXXX. Received YYYY; in original form ZZZZ}

\pubyear{\the\year{}}

\begin{document}
\label{firstpage}
\pagerange{\pageref{firstpage}--\pageref{lastpage}}
\maketitle

\begin{abstract}
Low-mass galaxies during the Epoch of Reionization are expected to form stars in short bursts separated by periods of inactivity, but what regulates this cycle, and how sensitive it is to the modelling of stellar feedback, remains unclear. We use the \textsc{megatron} high-redshift zoom simulation suite, in which four star-formation and stellar-feedback prescriptions are evolved from identical initial conditions, to study the regulation of stellar mass growth in galaxies with $M_* \sim 10^4 - 10^8$~M$_\odot$ at $z = 8.5$. We classify galaxies as Star-Forming, Mini-Quenched (no significant star formation over the last 10 Myr), or Quenched (no significant star formation over the last 100 Myr). The feedback prescription controls both how often galaxies are quiescent and how this depends on mass. Models with stronger or more clustered feedback yield larger quiescent fractions compared to the fiducial and Varying IMF models. Duty cycles at fixed mass also differ by factors of several between models. The cold gas profiles of Quenched galaxies are consistent with two physically distinct modes of quenching: mechanical gas evacuation, producing compact, cold gas-poor systems, and thermodynamic suppression, retaining a cold but star-formation-inactive reservoir. At fixed stellar mass, quiescent galaxies are substantially fainter than their star-forming counterparts, making the suppressed population particularly difficult to detect in unlensed JWST surveys and introducing a strong selection bias towards actively star-forming systems. Measuring the quiescent fraction as a function of stellar mass in lensing fields would provide a direct test of stellar-feedback models at Cosmic Dawn.
\end{abstract}



\section{Introduction}
\label{sec:introduction}

Low-mass galaxies at Cosmic Dawn are expected to experience highly time-variable star formation. Their shallow gravitational potentials make them particularly susceptible to stellar feedback, allowing individual episodes of star formation and supernova activity to heat, displace, or expel the cold gas required for subsequent star formation \citep{2005ApJ...624..726M, 2017MNRAS.466...11R, 2023MNRAS.521.2196A, 2023MNRAS.525.3806M, 2025MNRAS.541.1195R}. Rather than evolving smoothly along a continuous star-forming sequence, early dwarf galaxies may therefore cycle rapidly between bursts of star formation and intervals of strongly suppressed activity. Understanding what controls the incidence, duration, and recovery from these suppressed phases provides a sensitive test of stellar-feedback physics in the early Universe.

Throughout this work we use the term dwarf galaxy to refer broadly to systems with stellar masses $M_\ast \lesssim 10^9\,{\rm M_\odot}$. 
The resolved high-redshift population analysed here occupies the lower-mass regime $M_\ast \sim 10^4$--$10^8\,{\rm M_\odot}$. These systems are both abundant at high redshift and particularly sensitive to feedback because of their shallow gravitational potentials. Their response to stellar feedback therefore offers a stringent laboratory for testing how star formation is regulated during the Epoch of Reionization (EoR).

Observationally, \textit{JWST} is beginning to reveal that suppressed star formation is already present within the early galaxy population. While much of the initial work on high-redshift quiescence focused on comparatively massive systems at $z\sim3$--5 \citep[e.g.][]{2023Natur.619..716C,2024MNRAS.534..325C,2025ApJ...983...11W}, observations approaching the EoR indicate that inactivity in lower-mass galaxies can instead be short-lived and highly episodic.

Closer to the EoR, suppressed star formation need not represent sustained quenching. At $z=7.3$, \citet{2024Natur.629...53L} identified a low-mass galaxy whose recent star-formation history is consistent with a short burst followed by rapid, temporary suppression, motivating the term mini-quenched. Evidence for the subsequent rejuvenation of such systems has since been reported at $z\simeq7.9$, where the simultaneous presence of an evolved stellar population and renewed nebular emission indicates a return to active star formation following a short-lived quiescent phase \citep{2025MNRAS.537..112W}. Complementary searches for `smouldering' and dormant galaxies suggest that temporarily suppressed systems may constitute a non-negligible component of the low-mass galaxy population during the EoR \citep[e.g.][]{2025MNRAS.537.3662T, 2026A&A...705A.155C}.

Such systems may represent the low-SFR phases of intrinsically bursty star-formation histories rather than a population undergoing irreversible shutdown. Population-level measurements reinforce this interpretation. The increasing scatter about the star-forming main sequence, particularly when star formation is measured over short timescales and at low stellar masses, points towards rapid fluctuations between episodes of intense star formation and relative inactivity \citep[e.g.][]{2025ApJ...979..193C, 2025MNRAS.544.4551S}. High-resolution simulations similarly predict deviations above and below the star-forming main sequence, driven by cycles of starbursts, feedback-induced suppression, and subsequent recovery \citep[e.g.][]{2015MNRAS.451..839D, 2017MNRAS.466...88S, 2018MNRAS.480.4842C, 2025MNRAS.544.1732M}. Determining what controls these cycles, why some galaxies recover while others remain quenched, and how this behaviour depends on stellar mass and feedback physics therefore remains a fundamental open question in early galaxy evolution.

Stellar feedback in low-mass galaxies is inherently multi-channel, with photoionisation, stellar winds, and supernovae acting together to regulate the gas supply and star-formation cycle, while their effectiveness depends on the structure and thermodynamic state of the multiphase ISM
\citep{2022NatAs...6..647C}. Rather than attributing burstiness or temporary suppression to any single feedback channel, we therefore focus on their emergent impact on the star-formation history and duty cycle.

A growing body of high-redshift simulations and empirical models predicts that star formation becomes increasingly stochastic towards lower galaxy and halo masses and at earlier cosmic times. In the \textsc{serra} simulations, feedback-regulated low-mass galaxies repeatedly move between active and temporarily quiescent phases, with their star-forming duty cycle decreasing towards lower stellar mass \citep{Gelli_2023, 2025ApJ...985..126G}. High-redshift simulations similarly predict strong short-timescale variability driven by repeated cycles of gas inflow, star formation, feedback, and subsequent recovery \citep{2025MNRAS.544..513M, 2026arXiv260922431S}. Complementing these theoretical results, \citet{2026MNRAS.547ag415M} use UV luminosity functions, galaxy clustering, and H$\alpha$/UV ratios at $z\sim4$--6 to infer that star-formation variability increases systematically towards lower halo mass, reaching $\gtrsim1$ dex for $M_{\rm h}\lesssim10^9\,{\rm M_\odot}$, with a characteristic burst timescale of $\sim20$\,Myr. This empirical mass dependence is qualitatively consistent with the expectation that the shallow potential wells of low-mass systems make their star formation particularly susceptible to feedback-driven fluctuations. Together, these results emphasise that the duty cycle and duration of suppressed phases depend on the coupled evolution of gas supply, star formation, and multiple feedback processes, rather than on supernova energetics alone.

Despite this broad agreement that early star formation is highly time-variable, there is not yet consensus on either the magnitude of the burstiness or the abundance and duration of temporarily suppressed systems. Different theoretical approaches predict substantially different
star-forming duty cycles and quiescent fractions, while also differing in their treatment of star formation, stellar feedback, radiative transfer, resolution, and cosmological environment
\citep[e.g.][]{2025ApJ...985..126G, 2025MNRAS.544..513M}. Comparisons between independent simulation suites therefore cannot cleanly determine which differences arise specifically from the adopted star-formation and feedback prescriptions.

A controlled comparison is therefore needed to determine which properties of the early dwarf population are robust to the assumed feedback prescription and which are direct consequences of it. In particular, feedback may regulate not only the fraction of galaxies observed in a suppressed state, but also the duration and recurrence of those episodes, their dependence on galaxy mass, and the physical state of the gas from which subsequent star formation proceeds. Holding the underlying cosmological environment fixed provides a direct means of isolating these effects.

In this paper, we address these questions using the \textsc{megatron} high-redshift zoom simulation suite. \textsc{megatron} comprises four distinct simulations evolved from identical initial conditions and with the same numerical resolution and large-scale environment. Each run implements a different prescription for Population II star formation and stellar feedback. Because all four runs share identical initial conditions, the suite provides a controlled experiment for determining how these sub-grid choices alter the regulation of star formation in the same underlying high-redshift galaxy population.

Our principal aims are as follows:
\begin{enumerate}

    \item Quantify how different star-formation and stellar-feedback prescriptions control the absolute abundance, fractions, and global physical properties of quenched dwarf galaxies at $z\sim8.5$.

    \item Determine how the recent star-formation state of galaxies relates to their previous stellar-mass growth and star-forming duty cycle, and establish which of these relationships remain robust across the different feedback prescriptions.

    \item Characterise the physical mechanisms through which different feedback prescriptions suppress star formation, including their effects on the cold-gas reservoir and on the environments in which stars and supernovae form.

    \item Identify observable signatures of these differences, with particular emphasis on whether measurements of the suppressed fraction as a function of stellar mass can discriminate between feedback models.

\end{enumerate}

The remainder of this paper is organised as follows.
Section~\ref{sec:methods} describes the \textsc{megatron} simulation suite, the four sub-grid prescriptions, and our classification of star-forming and suppressed systems.
Section~\ref{sec:results} presents the incidence and mass dependence of suppressed star formation, the corresponding duty cycles and gas properties, and their observable signatures. Finally, Section~\ref{sec:discussion} discusses the implications for stellar-feedback models and the prospects for testing these predictions observationally.

\section{Methodology}
\label{sec:methods}

We analyse data from the high-redshift suite of the \textsc{megatron} cosmological simulation project \citep{ 2025arXiv251005201K}. The suite consists of four distinct, high-resolution radiation-hydrodynamical zoom simulations evolved from identical initial conditions (ICs) to $z = 8.5$, while varying the sub-grid physics governing Population~II (Pop~II) star formation efficiency, supernova energetics, and the stellar Initial Mass Function (IMF). This controlled configuration enables us to isolate the macroscopic effects of baryonic feedback physics from the underlying cosmological structure growth. 

Below, we summarise the core hydrodynamical and non-equilibrium radiative thermochemistry baseline, outline the four specialised sub-grid feedback prescriptions and define our quiescence criteria.

\subsection{The High-Redshift suite}
\label{sec:gal_model}

Each realization within the \textsc{megatron} suite is a cosmological zoom-in simulation featuring a dark matter particle mass resolution of $m_{\mathrm{DM}} = 2.5 \times 10^4\,\mathrm{M}_\odot$ in the high-resolution region.

The simulations use a constant-comoving refinement scheme that triggers cell splits via a quasi-Lagrangian overdensity threshold, while strictly enforcing that the local gas Jeans length is resolved by a minimum of four grid cells. Consequently, the maximum physical spatial resolution scales from $\Delta x \approx 1\,\mathrm{pc}$ at $z \sim 30$ when the first structures collapse, to $\Delta x \approx 5\,\mathrm{pc}$ at the target analysis redshift of $z = 8.5$ \citep{2025arXiv251005201K}.

The equations of gravity, hydrodynamics, and radiative transfer are solved self-consistently on-the-fly using the adaptive mesh refinement (AMR) code \textsc{ramses-rtz} \citep{2022MNRAS.512..348K}, which constitutes a specialized fork of \textsc{ramses} \citep{2002A&A...385..337T} and \textsc{ramses-rt} \citep{2013MNRAS.436.2188R}. Radiative transport, non-equilibrium thermochemistry, and the associated heating and cooling processes are coupled self-consistently through the \textsc{prism} framework \citep{2022arXiv221104626K, 2026OJAp....956097K}, tracking the precise ionization and chemical evolution of primordial species, molecules, and heavy elements across 70 distinct ionization states (including \ion{H}{i}--\ion{ii}, \ion{He}{i}--\ion{iii}, $\mathrm{e}^-$, \ion{C}{i}--\ion{vi}, \ion{N}{i}--\ion{vii}, \ion{O}{i}--\ion{viii}, \ion{Ne}{i}--\ion{x}, \ion{Mg}{i}--\ion{x}, \ion{Si}{i}--\ion{xi}, \ion{S}{i}--\ion{xi}, \ion{Fe}{i}--\ion{xi}, $\mathrm{H}_2$, and $\mathrm{CO}$). Higher ionization states not explicitly followed are assumed to maintain collisional ionization equilibrium. 

Star formation is governed by the local turbulent properties of the ISM \citep{2007ApJ...661..972P, 2012ApJ...761..156F, 2017MNRAS.466.4826K, 2018MNRAS.479..994R}. A gas cell becomes eligible for star formation only if it satisfies a strict set of thresholds: a local number density $n_{\mathrm{H}} > 10\,\mathrm{cm}^{-3}$, a localized density maximum, a locally convergent fluid flow ($\nabla \cdot \mathbf{v} < 0$), and a turbulent Jeans length smaller than the individual cell width. Cells meeting these criteria convert gas into stars via a Schmidt law \citep{1959ApJ...129..243S} with an efficiency per free-fall time ($\epsilon_{\mathrm{ff}}$) regulated by local Mach numbers and turbulence.

At low metallicities ($Z < 10^{-6}\,\mathrm{Z}_\odot$, where $\mathrm{Z}_\odot = 0.02$), star formation proceeds in Pop~III mode. Pop~III stars form as individual point-mass sources sampling a lognormal IMF with a characteristic mass of $100\,\mathrm{M}_\odot$ \citep{ 2014MNRAS.442.2560W, 2017MNRAS.466.4826K}. Their feedback pathways depend strictly on progenitor mass. Stars within the range $10 < m_\star \le 20\,\mathrm{M}_\odot$ undergo core-collapse supernovae (CCSNe; $E_{\mathrm{SN}} = 10^{51}\,\text{erg}$), those spanning $20 < m_\star \le 40\,\mathrm{M}_\odot$ terminate as hypernovae (HNe; $E_{\mathrm{HN}} \approx 2 \times 10^{52}\,\text{erg}$), and those in the range $140 \le m_\star \le 300\,\mathrm{M}_\odot$ undergo pair-instability supernovae (PISNe; $E_{\mathrm{PISN}} \approx 5 \times 10^{52}\,\text{erg}$, see exact mass-dependent scalings in \citealt{1959ApJ...129..243S}). Chemical yields and radiative inputs are implemented following \citet{2013ARA&A..51..457N} and \citet{2002A&A...382...28S}, respectively, while stars outside these specific mass intervals collapse directly into black holes without injecting kinetic energy or synthesised elements \citep[see structural details in][]{2026MNRAS.548ag529S}.

Once the gas crosses the metallicity threshold $Z \ge 10^{-6}\,\mathrm{Z}_\odot$, the simulation transitions to Pop~II star formation. Here, stellar particles represent simple stellar populations (SSPs) rather than individual stars, assuming a standard \citet{2001MNRAS.322..231K} IMF and a minimum initial stellar particle mass of $500\,\mathrm{M}_\odot$. These Pop~II stellar particles inject continuous stellar winds, execute supernova explosions, and emit radiative feedback following the approach of \citet{2020MNRAS.491.1656A}. Supernova feedback is deposited as thermal energy ($10^{51}\,\text{erg}$ per event) if the local cooling radius is resolved by the grid and is otherwise deposited as a directional momentum injection ($3 \times 10^5\,\mathrm{M}_\odot\,\mathrm{km}\,\mathrm{s}^{-1}$). Non-rotating CCSNe elemental yields are sourced from \citet{2018ApJS..237...13L}, and radiative feedback properties are computed dynamically using the \textsc{bpass} v2.2.1 binary stellar evolution tables \citep{2016MNRAS.456..485S}. This underlying Pop~II feedback budget has been thoroughly verified at $z=0$ for both Milky Way analogues \citep{2021MNRAS.503.5826A} and field dwarf systems \citep{2025MNRAS.541.1195R}.

\subsection{Variations in Sub-Grid Feedback Prescriptions}
\label{sec:subgrid}

\begin{figure*}
    \centering
    \includegraphics[width=\linewidth]{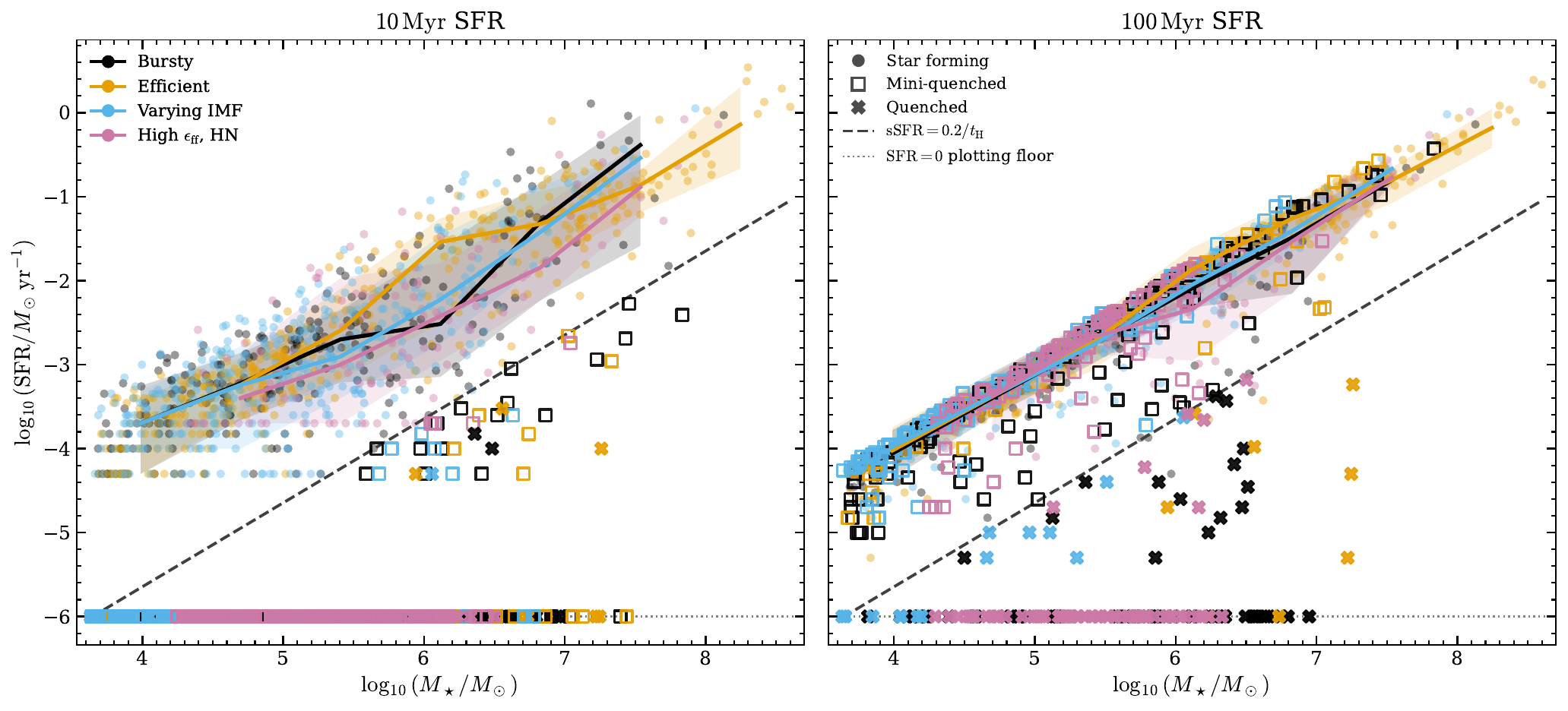}
    \caption{
    SFMS for the four \textsc{megatron} feedback prescriptions at $z=8.5$. The left and right panels show SFRs averaged over 10 and 100~Myr, respectively. Galaxies are classified as Star-Forming (circles), Mini-Quenched (open squares), or Quenched (crosses) according to the $\mathrm{sSFR}_{10}$ and $\mathrm{sSFR}_{100}$. Coloured solid curves show the median SFMS of the star-forming population in each simulation, while shaded regions denote the $1\sigma$ scatter. The horizontal dotted line marks the plotting floor adopted for systems with zero SFR over the corresponding averaging interval; their position at this floor should therefore not be interpreted as a measured SFR. The dashed diagonal line marks the quenching threshold, $\mathrm{SFR}=0.2M_\star/t_{\mathrm{H}}(z)$.
    }
    \label{fig:sfms_quenching}
\end{figure*}

We investigate the theoretical models proposed to reconcile the abundance of bright, high-redshift galaxies by considering four \textsc{megatron} realizations which systematically vary their Pop~II sub-grid and feedback modelling parameters:

\begin{itemize}
    \item \texttt{Efficient Star Formation (Efficient)}: This represents our fiducial baseline, utilising the standard Pop~II physics described in Section~\ref{sec:gal_model}. Because it features the lowest feedback energy per unit stellar mass among the configurations, it maintains the highest conversion efficiency of cold gas into stars, leading to highly compact and massive stellar cores.
    
    \item \texttt{Bursty Star Formation (Bursty)}: In this realization, the kinetic and thermal energy injected by all Pop~II supernova events is uniformly amplified by a factor of five ($5 \times 10^{51}\,\text{erg}$) relative to the fiducial run. This elevated feedback budget severely regulates total stellar mass growth and drives highly stochastic, bursty star formation histories characterized by massive, periodic gas clearings.
    
    \item \texttt{Varying IMF}: This model relaxes the assumption of a universal stellar birth cloud, allowing the high-mass slope of the Pop~II IMF to vary dynamically as a function of local gas density and metallicity following the relations of \citet{2012MNRAS.422.2246M}. In highly dense star-forming regions, the IMF becomes increasingly top-heavy, though it is capped at a maximum stellar mass of $120\,\mathrm{M}_\odot$. Radiative spectral energy distributions are extracted from \textsc{starburst99} \citep{1999ApJS..123....3L}, giving rise to stellar populations with exceptionally low mass-to-light ratios. Furthermore, a metallicity-dependent fraction of massive Pop~II stars undergo hypernova explosions \citep{2006ApJ...653.1145K} with varying explosion energies \citep{2006NuPhA.777..424N}.
    
    \item \texttt{High $\epsilon_{\mathrm{ff}}$, HN}: This model modifies both the conditions for star formation and the strength and clustering of stellar feedback. The star formation efficiency per free-fall time is fixed to a constant $\epsilon_{\mathrm{ff}} = 100\%$ rather than varying with local turbulence, and the minimum stellar particle mass is increased to $2000~\mathrm{M}_\odot$ to enforce highly clustered starbursts. In addition, the turbulent Jeans criterion is relaxed from $\lambda_{\rm J,turb}<\Delta x$ to $\lambda_{\rm J,turb}<4\Delta x$, allowing star formation to occur in lower-density gas than in the other models. Low-metallicity Pop~II stars are also permitted to undergo hypernova explosions identically to the \texttt{Varying IMF} run \citep{2006NuPhA.777..424N, 2006ApJ...653.1145K}. The modified star-formation criterion is primarily responsible for the characteristically lower-density ISM in this model, with strongly clustered star formation and energetic feedback further regulating its structure.
\end{itemize}

\subsection{Defining Quiescence}

We distinguish between active and quiescent galaxy populations within the  \textsc{megatron} simulation suite by adopting a redshift-dependent specific star formation rate ($\mathrm{sSFR}$) threshold. Rather than applying a fixed low-redshift cut, such as $\mathrm{sSFR} < 10^{-11}\,\mathrm{yr}^{-1}$, we define quiescence relative to the Hubble time at the redshift of interest. This is particularly important at high redshift, where galaxies have much shorter mass-doubling times and a fixed absolute threshold would not select quiescent systems consistently across cosmic time.

We therefore classify a galaxy as quenched when its $\mathrm{sSFR}$ falls below a fixed fraction of the inverse Hubble time, $1/t_{\mathrm{H}}(z)$. Specifically, we adopt the commonly used threshold
\begin{equation}
    \mathrm{sSFR} < \frac{0.2}{t_{\mathrm{H}}(z)},
\end{equation}
which has been widely used in studies of high-redshift quenched galaxies \citep[e.g.,][]{2008ApJ...688..770F, 2016ApJ...832...79P, 2020MNRAS.496..695C, 2026arXiv260115207C}.  We test the sensitivity of our results to this choice in Appendix~\ref{sec:quenching_threshold}, where we repeat the quenched-fraction analysis using thresholds of $0.1/t_{\mathrm{H}}(z)$ and $0.3/t_{\mathrm{H}}(z)$. These variations do not alter the qualitative ordering of the feedback models or our main conclusions.

At $z=8.5$, this corresponds to an absolute threshold of approximately $2.26 \times 10^{-10}\,\mathrm{yr}^{-1}$. This choice ensures that our selected galaxies have suppressed star formation relative to the rapidly assembling high-redshift star-forming main sequence, providing a robust definition for both transiently suppressed and quenched populations. Utilizing a physical metric like $\mathrm{sSFR}$ also bypasses the limitations inherent to observational frameworks. Conversely, colour-based quiescent selections can be incomplete and contaminated, since dusty star-forming galaxies or emission-line sources may enter passive colour regions \citep{2018A&A...618A..85S}, while some truly quiescent objects may fall outside conventional colour cuts \citep{ 2018MNRAS.473.2098M, 2025MNRAS.539..557B}.

We capture the episodic nature of high-redshift star formation and isolate the structural impacts of our various feedback models by evaluating the resolved galaxy population at $z=8.5$, restricted to systems containing $>1000$ gas cells and $>10$ stellar particles, across two distinct averaging timescales. We implement a three-way, mutually exclusive classification scheme based on specific star formation rates calculated over short-term ($\tau = 10\,\mathrm{Myr}$; $\mathrm{sSFR}_{10}$) and long-term ($\tau = 100\,\mathrm{Myr}$; $\mathrm{sSFR}_{100}$) intervals:
\begin{enumerate}
    \item \textit{Star-Forming (SF):} Systems actively maintaining ongoing star formation, defined by a short-term rate above the quenching limit ($\mathrm{sSFR}_{10} \ge 0.2/t_{\mathrm{H}}(z)$).
    \item \textit{Mini-Quenched (MiniQ):} Systems undergoing temporary or transient suppression, where the short-term rate drops below the threshold ($\mathrm{sSFR}_{10} < 0.2/t_{\mathrm{H}}(z)$) but the long-term average remains active ($\mathrm{sSFR}_{100} \ge 0.2/t_{\mathrm{H}}(z)$).
    \item \textit{Quenched:} Systems exhibiting long-term, sustained cessation of Star-Forming activity, where both timescales fall below the threshold ($\mathrm{sSFR}_{10} < 0.2/t_{\mathrm{H}}(z)$ and $\mathrm{sSFR}_{100} < 0.2/t_{\mathrm{H}}(z)$).
\end{enumerate}

This dual-timescale classification is illustrated in Fig.~\ref{fig:sfms_quenching}, where we show the location of the resolved galaxy populations in the stellar mass--SFR plane using SFRs averaged over 10 and 100~Myr, where solid curves trace the median star-forming main sequence (SFMS) of each feedback prescription. Galaxies classified as MiniQ lie below the quenching threshold on the 10~Myr timescale while retaining significant star formation when averaged over 100~Myr, distinguishing recent suppression from sustained quiescence. In contrast, Quenched galaxies remain suppressed on both timescales. Systems with no star formation over the corresponding averaging interval are shown at a fixed plotting floor. We note that this classification refers to the star-formation state at the epoch of observation and does not imply that star formation cannot subsequently rejuvenate.

\section{Results}
\label{sec:results}

\subsection{Quenched Galaxy Fractions Across Feedback Prescriptions}
\label{subsec:quenched_fractions}
 
\begin{figure}
    \includegraphics[width=\columnwidth]{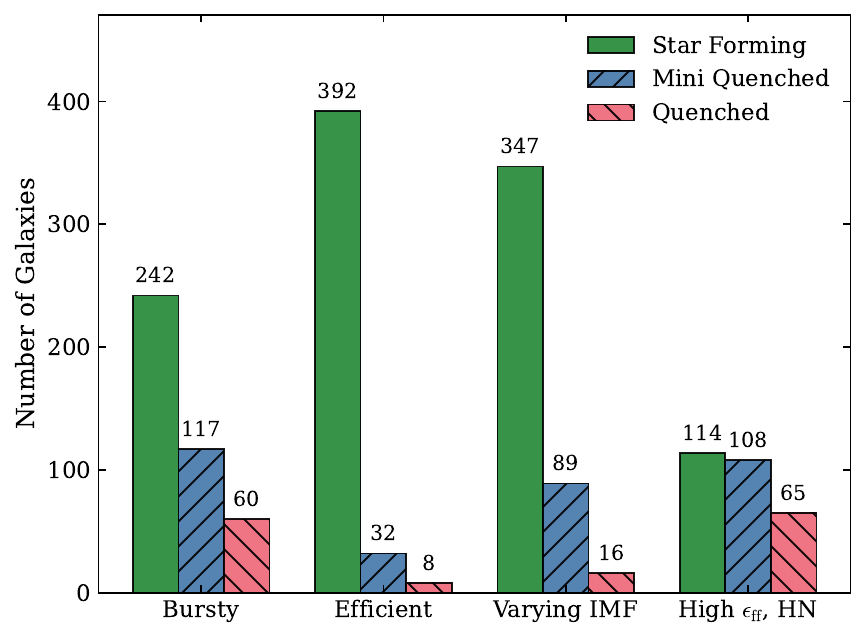}
    \caption{The absolute number counts of SF, MiniQ, and Quenched galaxies across the four \textsc{megatron} feedback prescriptions at $z = 8.5$. Only systems resolved with $>1000$ gas cells and $>10$ Pop~II stellar particles are included.}
    \label{fig:quenched_numbers}
\end{figure}
 
In Fig.~\ref{fig:quenched_numbers} we present the absolute numbers of MiniQ and Quenched galaxies identified at $z = 8.5$ for each of the four \textsc{megatron} feedback configurations. The \texttt{High $\epsilon_{\mathrm{ff}}$, HN} prescription yields the largest Quenched population, with 65~Quenched galaxies and 108~MiniQ systems. The \texttt{Bursty} model produces a comparable number of Quenched galaxies ($60$) with a marginally larger MiniQ count ($117$), followed by the \texttt{Varying IMF} model ($89$~MiniQ, $16$~Quenched) and the \texttt{Efficient} model ($32$~MiniQ, $8$~Quenched).  The differences in Quenched abundance indicate that the feedback prescription strongly regulates the ability of galaxies to maintain suppressed star formation over extended timescales, rather than merely producing short-lived interruptions to star formation.

The elevated supernova energy of the \texttt{Bursty} model ($5 \times 10^{51}$~erg) and the concentrated, hypernova-rich bursts of the \texttt{High $\epsilon_{\mathrm{ff}}$, HN} model are expected to generate highly episodic feedback events capable of disrupting the cold ISM and thereby increasing the likelihood of long-lived suppression of star formation in low-mass systems. By contrast, feedback in the \texttt{Efficient} model is deposited predominantly in denser gas, where it is less effective at accelerating and displacing the resolved ISM \citep{2025MNRAS.541.1195R, 2026OJAp....958199C}. These differences are examined further in Sec.~\ref{subsec:sf_sne_environments}.
 
\begin{figure*}
    \centering
    \includegraphics[width=\linewidth]{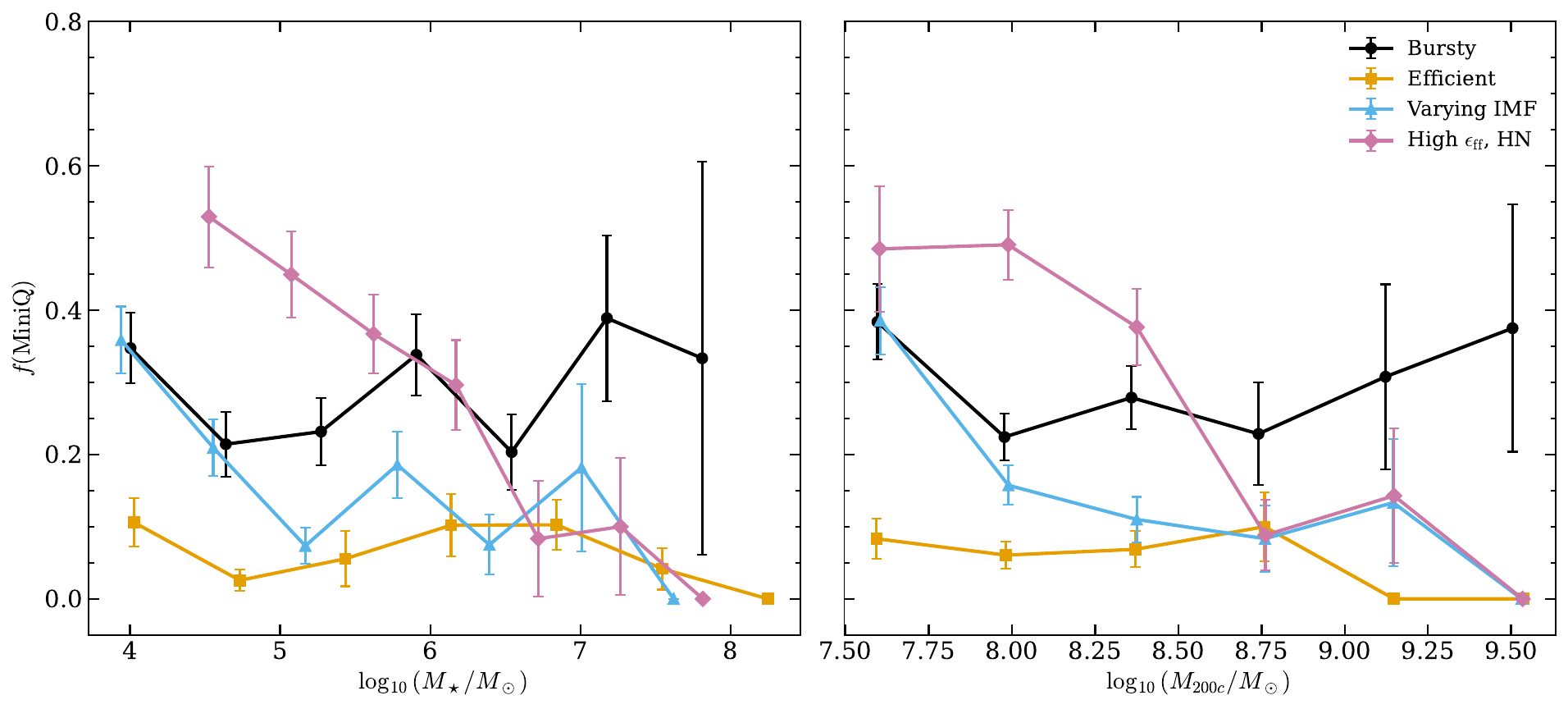}
    \caption{The fraction of MiniQ galaxies ($f_{\mathrm{MiniQ}}$) as a function of stellar mass (left) and halo mass (right)
             across the four feedback prescriptions. Error bars correspond to the binomial error in each bin. Mass bins are defined independently per prescription; only overlapping mass ranges should be compared across
             models.}
    \label{fig:quenched_fraction_mini}
\end{figure*}
 
\begin{figure*}
    \centering
    \includegraphics[width=\linewidth]{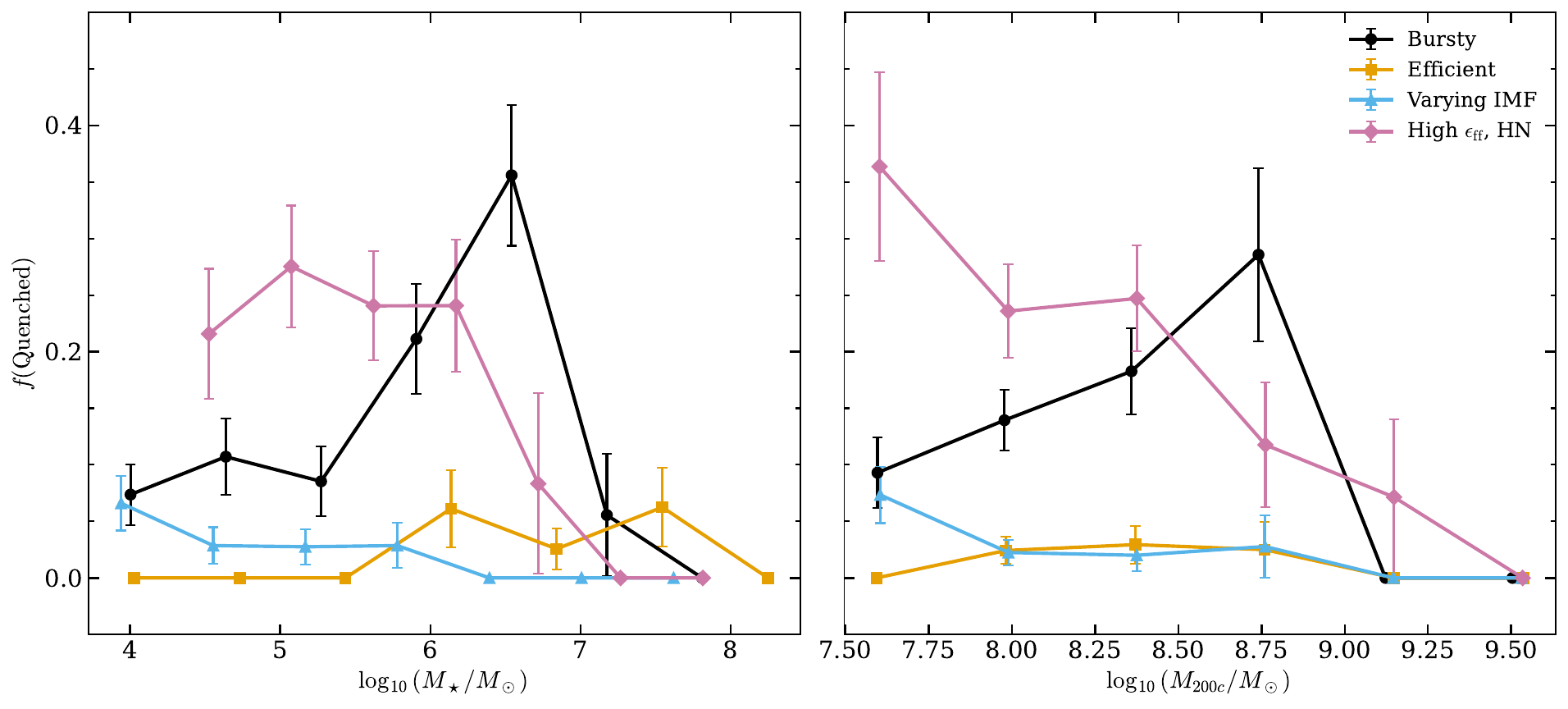}
    \caption{The fraction of Quenched galaxies ($f_{\mathrm{Quenched}}$) as a function of stellar mass (left) and halo mass (right)
             across the four feedback prescriptions. Binomial error bars as in Fig.~\ref{fig:quenched_fraction_mini}.}
    \label{fig:quenched_fraction_perm}
\end{figure*}
 
Figs.~\ref{fig:quenched_fraction_mini} and~\ref{fig:quenched_fraction_perm}
present the differential fractions of MiniQ and Quenched galaxies as functions of stellar mass, $M_{\star}$, and halo mass, $M_{200\mathrm{c}}$. Uncertainties correspond to the binomial standard error in each bin. Because mass bins are defined independently for each prescription, cross-model comparisons should be made within overlapping mass ranges rather than between bins of the same index. The highest-mass bins generally contain only a small number of systems ($\lesssim 12$) and are therefore statistically uncertain; we flag these where relevant.
 
When examining $f_{\mathrm{MiniQ}}$ as a function of $M_\star$
(Fig.~\ref{fig:quenched_fraction_mini}, left panel), the \texttt{High $\epsilon_{\mathrm{ff}}$, HN} model exhibits the strongest mass dependence, with $f_{\mathrm{MiniQ}}$ declining systematically from $0.529 \pm 0.070$ at $\log_{10}(M_{\star}/\mathrm{M}_{\odot}) \simeq 4.53$ to $0.296 \pm 0.062$ at $\log_{10}(M_{\star}/\mathrm{M}_{\odot}) \simeq 6.17$. This decline is consistent with deeper potential wells making feedback-driven interruptions of star formation less frequent. A consistent decline with halo mass is recovered in the right panel.

\begin{figure*}
    \centering
    \includegraphics[width=\linewidth]{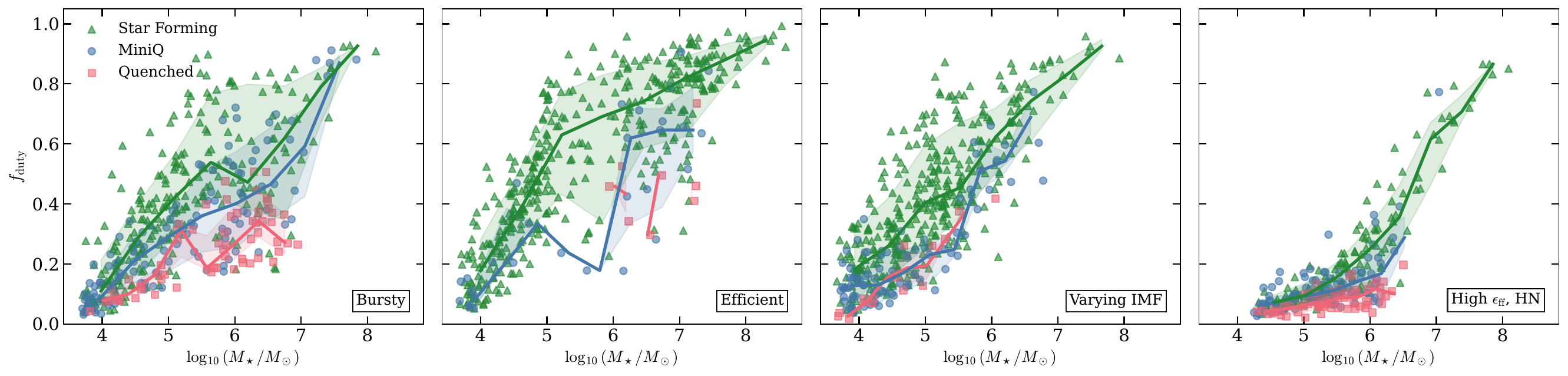}
    \caption{The duty cycle ($f_{\mathrm{duty}}$) against the stellar mass ($M_*$) for Star-Forming, MiniQ and Quenched galaxies in the four different models. The solid, colour-matched curves track the running median of $f_{\rm duty}$ across the stellar mass bins, while the surrounding lightly shaded bands mark the corresponding $1\sigma$ interval.
}
    \label{fig:duty_cycle}
\end{figure*}

The \texttt{Efficient} model produces a flat evolution of $f_{\mathrm{MiniQ}}$ with increasing $M_\star$, with no systematic trends, and substantially lower MiniQ fractions than the \texttt{Bursty} and \texttt{High $\epsilon_{\mathrm{ff}}$, HN} models for low stellar masses. Its halo-mass dependence is equally flat, remaining near $f_{\mathrm{MiniQ}} \simeq 0.06$--$0.10$ over $7.4 \lesssim \log_{10}(M_{200\mathrm{c}}/\mathrm{M}_{\odot}) \lesssim 9.0$. Feedback in this model therefore rarely interrupts star formation for $\gtrsim10$\,Myr at any mass.
 
The \texttt{Varying IMF} model displays MiniQ fractions that lie above \texttt{Efficient} but generally below \texttt{High $\epsilon_{\mathrm{ff}}$, HN} for $M_\star<10^{6.5}\,\mathrm{M}_\odot$ (matching \texttt{Bursty} only in the lowest two mass bins), and becomes similar to the \texttt{High $\epsilon_{\mathrm{ff}}$, HN} and \texttt{Efficient} models for higher $M_\star$, with the largest value of $0.358 \pm 0.047$ at $\log_{10}(M_{\star}/\mathrm{M}_{\odot}) \simeq 3.94$, declining to $0.073 \pm 0.025$ by the third mass bin. The elevated MiniQ fraction at the lowest masses may reflect the onset of hypernova feedback in galaxies that have accumulated sufficient stellar mass to host the first generation of massive, low-metallicity stars.

In contrast to the above models, the \texttt{Bursty} prescription maintains an elevated $f_{\mathrm{MiniQ}}$ over a broad mass interval, with values ranging between $\simeq 0.20$ and $0.39$ for $3.7 \lesssim \log_{10}(M_{\star}/\mathrm{M}_{\odot}) \lesssim 7.5$. This broad plateau is consistent with the amplified supernova energy ($5\times10^{51}$\,erg per event) remaining able to temporarily suppress star formation across a wider range of galaxy masses than in the other models. The halo-mass relation shows a comparable plateau.
 
The Quenched fractions shown in Fig.~\ref{fig:quenched_fraction_perm} reveal a different ordering of the models compared to the MiniQ fractions above. The \texttt{High $\epsilon_{\mathrm{ff}}$, HN} model produces the largest Quenched fractions at low and intermediate masses, with $f_{\mathrm{Quenched}}$ remaining approximately constant at $0.22$--$0.28$ over $4.3 \lesssim \log_{10}(M_{\star}/\mathrm{M}_{\odot}) \lesssim 6.4$ before declining at higher masses. The halo-mass relation shows the same trend: $f_{\mathrm{Quenched}}$ decreases from $0.364 \pm 0.084$ at $\log_{10}(M_{200\mathrm{c}}/\mathrm{M}_{\odot}) \simeq 7.60$ to $0.118 \pm 0.055$ at $\simeq 8.76$. The broad plateau in the \texttt{High $\epsilon_{\mathrm{ff}}$, HN} Quenched fraction, combined with the steep decline in $f_{\mathrm{MiniQ}}$ with increasing mass discussed above, is consistent with the enforced $\epsilon_{\mathrm{ff}}=100\%$ prescription concentrating stellar birth into highly clustered, explosive events that can clear the ISM of low-mass systems.

The \texttt{Bursty} model exhibits a qualitatively distinct, strongly mass-dependent $f_{\mathrm{Quenched}}$, with a localised enhancement at intermediate masses. The Quenched fraction rises from $0.074 \pm 0.027$ at $\log_{10}(M_{\star}/\mathrm{M}_{\odot}) \simeq 4.00$ to $0.211 \pm 0.048$ at $\simeq 5.91$, reaching a pronounced peak of $0.356 \pm 0.062$ at $\log_{10}(M_{\star}/\mathrm{M}_{\odot}) \simeq 6.54$.

This characteristic mass scale likely reflects a balance between the cumulative energy available from successive supernova generations and the increasing gravitational binding energy of the host halo. At lower stellar masses the stellar populations are small, so the energy injected by supernovae is insufficient to unbind the gas; it is either retained within the halo or rapidly re-accreted, and any suppression of star formation remains transient. Around $M_\star\sim10^{6.5}\,\mathrm{M}_\odot$, the amplified supernova budget can drive coherent, galaxy-wide outflows that evacuate the cold gas. Towards higher stellar masses, the increasing halo binding energy makes gas evacuation and delayed re-accretion progressively more difficult, so $f_{\mathrm{Quenched}}$ declines above the peak.

At the low-mass end, the \texttt{Efficient} model cannot maintain sustained quiescence at these masses. No Quenched galaxies are found below $\log_{10}(M_\star/\mathrm{M}_\odot)\simeq5.8$, despite these bins containing between $36$ and $117$ resolved systems. This complete absence reflects a fundamental transition in feedback efficiency. In the shallowest potential wells ($M_\star\lesssim10^{5.8}~\mathrm{M}_\odot$), the low-energy supernova injection is either retained or rapidly re-accreted, ensuring that any suppression of star formation remains transient. In addition, the \texttt{Efficient} model experiences the slowest progression of reionization among the four \textsc{megatron} simulations \citep[Appendix~C]{2025arXiv251005232R}. Although this run produces the largest number of ionizing photons owing to its elevated star formation rates \citep{2025arXiv251005201K}, its dense ISM results in a substantially lower escape fraction, delaying the build-up of the ionizing background \citep[][\textcolor{blue}{Choustikov et al., \textit{in prep.}}]{2026OJAp....958199C}. Consequently, the external photoionization heating that suppresses gas accretion onto the lowest-mass haloes is also weaker, likely further reducing the efficiency of sustained quenching in these systems. Only above $\log_{10}(M_\star/\mathrm{M}_\odot)\simeq6.1$ does the cumulative energy from successive supernova generations become sufficient to drive the large-scale outflows associated with sustained quenching, at which point $f_{\mathrm{Quenched}}$ plateaus at a modest $\simeq3$--$6$ per cent.
 
The \texttt{Varying IMF} model similarly produces very few Quenched systems, with $f_{\mathrm{Quenched}} \lesssim 0.07$ across all well-populated bins, indicating that suppression of star formation in this prescription is also predominantly short-lived.
 
Taken together, the MiniQ and Quenched fractions demonstrate that the feedback prescription influences not only the incidence of suppressed star formation but also its duration. The \texttt{High $\epsilon_{\mathrm{ff}}$, HN} model is associated with the highest Quenched fractions at low stellar masses, suggesting that its clustered, $\epsilon_{\mathrm{ff}}=100\%$ starbursts may convert temporary suppressions into inactive phases lasting $\gtrsim100$\,Myr more efficiently. The \texttt{Bursty} model sustains frequent temporary suppression across a wider mass range, with a distinct peak in $f_{\mathrm{Quenched}}$ at $M_\star \approx 10^{6.5}\,\mathrm{M}_\odot$. By contrast, the \texttt{Efficient} and \texttt{Varying IMF} prescriptions are far more likely to interrupt star formation transiently than to keep it suppressed for $\gtrsim100$\,Myr, underscoring the sensitivity of early dwarf galaxy star formation histories to both the energy scale and clustering of stellar feedback.

\subsection{Duty Cycles as Discriminators of Feedback Physics}
\label{subsec:duty_cycle}

 We next examine the star-formation duty cycle, $f_{\rm duty}$, which encodes information about both the total active time and the episodic nature of star formation within each model so we can further differentiate the four \textsc{megatron} feedback prescriptions. This provides a diagnostic that is highly complementary to integrated quantities such as stellar mass, especially when characterizing dwarf systems sensitive to rapid structural variations \citep{Gelli_2023}. Following the parametric formulation formalized by \citet{Gelli_2023} to evaluate early low-mass systems, the duty cycle is defined as:
\begin{equation}
f_{\rm duty}=\frac{t_{\rm on}}{t_{\rm obs}-t_{\rm form}}
\label{eq:fduty}
\end{equation}

where $t_{\rm on}$ is the total time during which a galaxy was actively forming stars (${\rm SFR}_{1\,\mathrm{Myr}}>0$), $t_{\rm obs}$ is the cosmic time at the epoch of observation ($z=8.5$), and $t_{\rm form}$ is the cosmic time at which the first surviving stellar particle formed in the galaxy. Representative $1\,\mathrm{Myr}$-binned star-formation histories for the three most massive galaxies in each model and star-formation class are shown in Appendix~\ref{sec:representative_sfhs}, providing a direct visualisation of the active and inactive intervals that enter the calculation of $t_{\rm on}$ and hence $f_{\rm duty}$. This metric is particularly suited to diagnosing the highly stochastic, ``bursty'' star-formation modes predicted at cosmic dawn \citep{2024MNRAS.527.2139D,2025MNRAS.544..513M} and evaluating whether early quenched systems are undergoing sustained or transient periods of quiescence.

In Fig.~\ref{fig:duty_cycle}, we show the distribution of individual galaxy duty cycles as a function of total stellar mass, $\log_{10}(M_\star / M_\odot)$, across the entire resolved population, providing a visual measure of the statistical dispersion within each sub-population.

A universal signature emerges across all configurations: quenched systems consistently display substantially lower duty cycles than their active counterparts at equivalent stellar masses. This systematic divergence confirms that the path to quenching fundamentally alters the manner and efficiency with which dwarf galaxies accumulate stellar mass, regardless of the specific code implementation. 

However, the particular behaviour of $f_{\rm duty}$ across the models reveals fundamentally distinct physical modes of regulation. Within the \texttt{High $\epsilon_{\mathrm{ff}}$, HN} model, Quenched galaxies exhibit remarkably low and flat duty cycles, tracking from a median of $f_{\rm duty} = 0.043$ at $M_{\star} \sim 10^{4.5}~M_{\odot}$ up to only $f_{\rm duty} = 0.110$ at $M_{\star} \sim 10^{6.4}~M_{\odot}$. This suppressed duty cycle is shared by the active Star-Forming population, which exhibits a median $f_{\rm duty} \sim 0.07$--$0.35$ below $M_{\star} = 10^{6.5}~M_{\odot}$. This points to a highly stochastic, cyclical mode of growth; highly clustered starbursts ($\epsilon_{\rm ff}=100\%$, $2000~M_\odot$ minimum stellar particle mass), aided by hypernovae, trigger violent feedback events that rapidly evacuate gas from shallow potential wells, forcing long dynamical delays for cooling and re-accretion. Systems that enter the Quenched population in this framework are those whose early starbursts are sufficiently energetic to strongly deplete their gas reservoirs, or those that experience early environmental starvation following their first star-forming cycle. This pervasive burstiness creates a notable observational caveat: at these high redshifts, distinguishing active dwarfs from genuinely quenched systems based solely on instantaneous $\mathrm{SFR}$ indicators will be exceedingly challenging, as even active galaxies spend the vast majority of their lifetimes in a dormant state.

In sharp contrast, the \texttt{Efficient} feedback prescription yields an entirely different evolutionary pathway. In this model, galaxies retain high median duty cycles ($f_{\rm duty} \sim 0.34$--$0.49$). This elevated baseline implies a continuous, monolithic star formation mode where feedback does not drive frequent, short-term gas expulsions. Galaxies in the \texttt{Efficient} model grow steadily and uninterruptedly; quenching occurs exclusively as a late-stage, abrupt event when a galaxy has already accumulated significant active time ($t_{\rm on}$) prior to its final shutdown.

The \texttt{Varying IMF} and \texttt{Bursty} models occupy intermediate duty cycles, but with different stellar-mass dependences. For the \texttt{Varying IMF} model, the median duty cycle of Quenched galaxies scales smoothly from $0.037$ at $M_{\star} \sim 10^{4}~M_{\odot}$ to $0.418$ at $M_{\star} \sim 10^{6.1}~M_{\odot}$. This steep gradient reflects the changing energy budget of the stellar populations: at low metallicities and early times, a top-heavy IMF yields an excess of massive stars, generating intense supernova feedback per unit mass that suppresses early duty cycles. As chemical enrichment proceeds, the IMF normalizes, feedback leverage diminishes, and star formation transitions into a more continuous, high-duty-cycle mode. 

The \texttt{Bursty} model shows an overall increase in the duty cycle of Quenched galaxies with stellar mass, but the relation is shallower and less monotonic than that of the Star-Forming population. At $M_\star \gtrsim 10^{5.5}~M_\odot$, the Quenched median fluctuates around $f_{\rm duty}\sim0.2$--$0.35$, rather than continuing to rise steadily towards the high values reached by actively Star-Forming systems. Thus, even relatively massive Quenched galaxies have formed stars during only a minority of the time since their first star-formation event. This behaviour is consistent with the amplified supernova feedback in this model ($5\times10^{51}$\,erg per event) maintaining substantial periods of inactivity despite the greater resistance to gas removal expected in deeper potential wells.

Finally, the MiniQ populations across all models occupy a well-defined intermediate regime that supports their interpretation as transiently suppressed systems. In the \texttt{High $\epsilon_{\mathrm{ff}}$, HN} configuration, MiniQ galaxies steadily step up from $f_{\rm duty} = 0.062$ ($\sim 10^{4.5}~M_{\odot}$) to $0.288$ ($\sim 10^{6.4}~M_{\odot}$). At all masses, they track noticeably below the active main sequence but remain significantly above the absolute floor set by the Quenched population, highlighting their ongoing capacity to cool gas and resume star formation after a temporary quiescent phase. This behavior is mirrored in the \texttt{Efficient} model, where MiniQ galaxies reach high medians of $f_{\rm duty} \sim 0.53$--$0.66$ at $M_{\star} \sim 10^{6.1}$--$10^{7.4}~M_{\odot}$, confirming that transient quiescence in that framework occurs only after extensive, continuous mass assembly. This population mimics the properties of temporarily quiescent galaxies currently being uncovered at high redshifts by JWST observations \citep[e.g.,][]{2023ApJ...949L..23S, 2024Natur.629...53L, 2025ApJ...985..126G}. We examine this connection in greater detail in Sec.~\ref{subsec:observational_signatures}, where we compare the spectral properties of the simulated SF, MiniQ, and Quenched populations using their UV slopes, Balmer-break strengths, and full spectral energy distributions.

\subsection{Stellar mass assembly}
\label{subsec:mass_assembly}

Reconstructed stellar-mass assembly histories are increasingly being used to characterise the growth of high-redshift galaxies and to compare observed systems with theoretical models. For example, \citet{2026arXiv260513966C} infer the lookback times at which galaxies assembled different fractions of their stellar mass from \textit{JWST}/NIRSpec spectroscopy. While comparisons between such reconstructed histories and simulations are becoming an important test of early galaxy formation, these quantities are, however, sensitive both to the assumed SFH reconstruction \citep{2025MNRAS.537.1826T,2026arXiv260521599D} and to how the galaxy sample itself is selected.

Our sSFR-based classification provides a direct example of the latter effect. In the raw comparison, Quenched galaxies assemble 25, 50, and 75 per cent of their final stellar mass earlier than MiniQ and Star-Forming systems at fixed stellar mass. However, the Quenched criterion explicitly requires little recent star formation over the preceding 100\,Myr, so part of this apparent early assembly is expected by construction. We test this by removing the final 100\,Myr of star formation from Star-Forming SFHs and matching the resulting sample in stellar mass and feedback-model composition to the Quenched population. The full comparison is presented in Appendix~\ref{sec:assembly_selection_test}.

Once the same inactivity window is imposed, most of the separation disappears. In particular, the differences in $\tau_{50}$ and $\tau_{75}$ are consistent with the scatter of the matched Star-Forming population, while the Quenched systems do not show an earlier $\tau_{25}$ than the null expectation. We therefore find no evidence that Quenched galaxies follow a distinct, systematically earlier stellar-mass assembly pathway beyond that induced by their lack of recent star formation. This provides a useful caution for observational comparisons: differences in reconstructed assembly times should be interpreted together with the recent-SFR criteria used to define the samples.

While the integrated assembly times are therefore largely insensitive to the feedback prescription, the duration and completeness of the suppressed phase are not. We quantify this using the time since the most recent non-zero star formation in the 1\,Myr-averaged SFH, $t_{\rm last}$.

\begin{figure}
    \centering
    \includegraphics[width=\columnwidth]{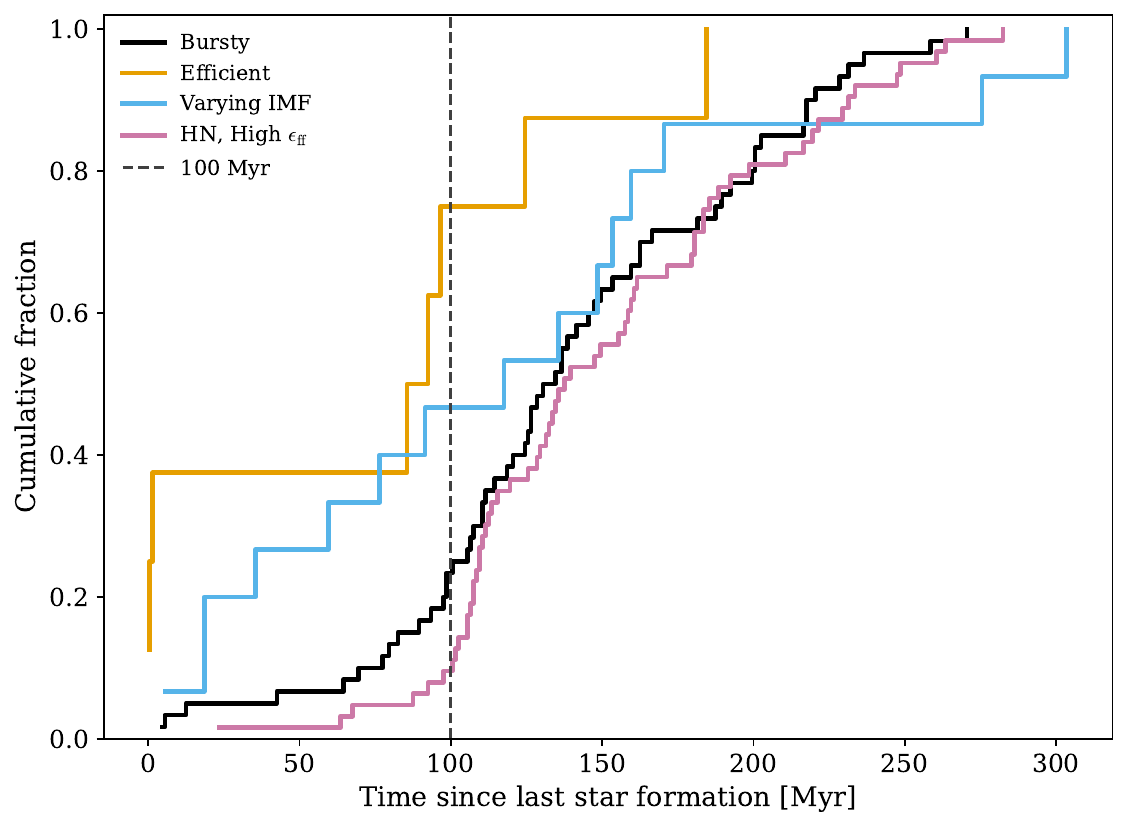}
    \caption{
    Cumulative distribution of the time since the most recent non-zero star formation in the 1-Myr SFH, $t_{\rm last}$, for Quenched galaxies in each feedback prescription. The dashed vertical line marks 100\,Myr, the averaging interval of our long-timescale quiescence criterion. 
    }
    \label{fig:tlast}
\end{figure}

Fig.~\ref{fig:tlast} shows that the $t_{\rm last}$ distributions differ substantially between the four feedback prescriptions. In the \texttt{Bursty} and \texttt{High~$\epsilon_{\rm ff}$, HN} models, 77 and 89 per cent of Quenched galaxies, respectively, have formed no stars during the preceding 100\,Myr, with median $t_{\rm last}=133$ and $138$\,Myr. In contrast, only 25 per cent of Quenched galaxies in the \texttt{Efficient} model and 56 per cent in the \texttt{Varying IMF} model have $t_{\rm last}>100$\,Myr; the remainder continue to form stars at a sufficiently low level to remain below the 100-Myr sSFR threshold. The small Quenched samples in the \texttt{Efficient} and \texttt{Varying IMF} prescriptions warrant caution when interpreting these fractions.

This distinction anticipates the gas-phase results of Sec.~\ref{subsec:sf_sne_environments}. Models in which suppression is associated with strong depletion of the cold reservoir preferentially produce genuinely dormant systems, whereas models that retain compact cold gas more often permit residual, low-level star formation. The feedback prescription therefore regulates the character and persistence of the suppressed state, how star formation is interrupted, how completely it ceases, and how long that interruption is maintained.

\subsection{Physical Drivers: Feedback Coupling and Gas Phase Structure}
\label{subsec:sf_sne_environments}
\begin{figure*}
    \centering
    \includegraphics[width=\linewidth]{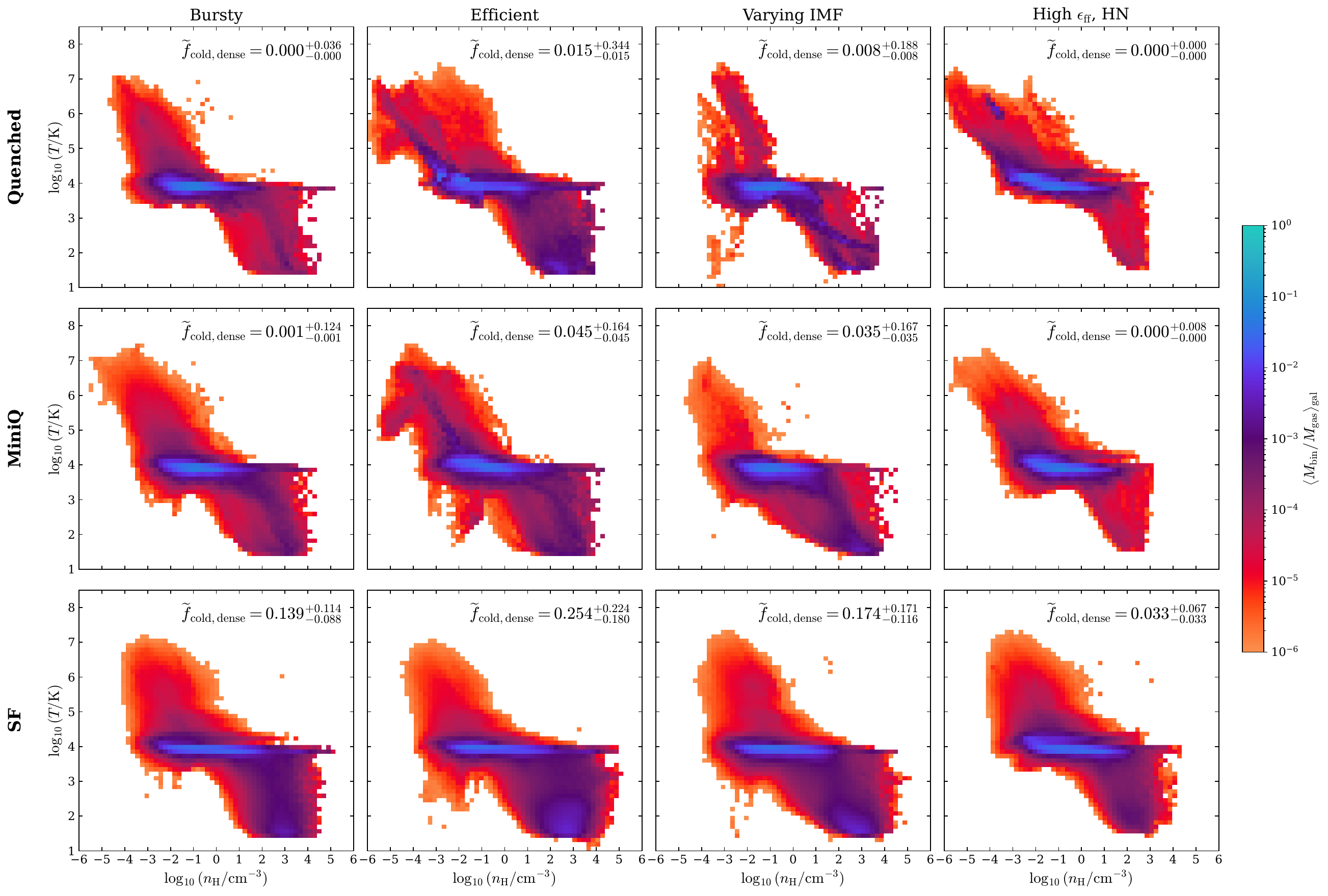}
    \caption{Population-stacked gas temperature--density phase diagrams for the four \textsc{megatron} feedback prescriptions at $z=8.5$. Rows show Quenched, MiniQ, and actively SF galaxies, while columns correspond to the four feedback models. For each galaxy, the gas mass in each temperature--density bin is divided by that galaxy's total gas mass, $M_{\rm bin}/M_{\rm gas}$, such that each galaxy contributes a normalised gas mass distribution. These distributions are then averaged across all galaxies in the corresponding population. The annotation in the upper-right corner of each panel gives the median fraction of gas in the cold, dense phase, $\widetilde{f}_{\rm cold, dense}$, defined by $T<10^{3.5}\,\mathrm{K}$ and $n_{\rm H}>10\,\mathrm{cm}^{-3}$, with the $1\sigma$ scatter across galaxies in the corresponding stack.
    }
    \label{fig:gas_phase_diagram}
\end{figure*}
To investigate the physical origin of the differences in quenching efficiency between the feedback prescriptions, we examine the thermodynamic structure of the gas within the \textsc{megatron} suite. The effectiveness of stellar feedback depends on both the injected energy and the environment in which it is deposited. Feedback acting in lower-density gas can experience reduced radiative losses and couple more effectively to the surrounding ISM, favouring the development of outflows.

Within the same simulation suite, \citet{2026OJAp....958199C} show that the \texttt{Efficient} model produces the densest ISM, which they attribute to its weaker feedback. In the \texttt{High $\epsilon_{\rm ff}$, HN} model, stars preferentially form in lower-density environments, enabling subsequent supernovae to couple more effectively to the gas; hypernovae additionally assist in clearing stellar birth clouds. These results establish that both feedback strength and ambient gas density contribute to the differences in ISM structure between the models.

Motivated by this interpretation, we investigate whether quenched galaxies retain cold, dense gas or instead exhibit a depletion of this component. Fig.~\ref{fig:gas_phase_diagram} presents the mean gas mass fraction in the temperature--density plane for each population--model combination. Each galaxy is first normalised by its total gas mass before the resulting phase distributions are averaged, such that individual gas-rich systems do not dominate the population stack. The accompanying cold, dense gas fraction, $\widetilde{f}_{\rm cold,dense}$, provides a complementary galaxy-level measure of the availability of gas approaching star-forming conditions.

The \texttt{Bursty} model exhibits a clear depletion of the cold, dense gas phase with increasing quenching severity. The actively SF population contains a substantial cold, dense component, with $\widetilde{f}_{\rm cold, dense}=0.139^{+0.114}_{-0.088}$, whereas the median falls to $\widetilde{f}_{\rm cold, dense}=0.001^{+0.121}_{-0.001}$ in the MiniQ galaxies and zero in the Quenched populations. The phase diagrams similarly show that gas at $n_{\rm H}>10\,{\rm cm}^{-3}$ and $T<10^{3.5}\,{\rm K}$ becomes progressively depleted towards the quenched populations. The remaining gas in Quenched systems is instead dominated by a diffuse warm-to-hot component. Together, these trends are consistent with gas that has been mechanically heated and displaced by the amplified supernova budget.

A similar behaviour is seen in the \texttt{High $\epsilon_{\rm ff}$, HN} model. Its actively SF galaxies contain a measurable cold, dense component, with $\widetilde{f}_{\rm cold,dense}=0.033^{+0.067}_{-0.033}$, which falls to a median of zero in MiniQ and Quenched populations. This is consistent with a mechanical blowout interpretation, in which highly concentrated hypernova feedback completely evacuates the central ISM, and is not an artefact of a single extreme system but holds as a population-wide outcome. Even in the MiniQ population of this model, cold dense gas is already strongly reduced relative to the active population, indicating that these systems are caught in the process of ISM evacuation rather than in a thermodynamically suppressed state.

The \texttt{Efficient} and \texttt{Varying IMF} models present a qualitatively different picture. Their actively SF populations possess particularly large cold, dense gas fractions, with $\widetilde{f}_{\rm cold, dense}=0.254^{+0.224}_{-0.180}$ and $0.174^{+0.171}_{-0.116}$, respectively. Although these fractions decline substantially in the MiniQ populations, to $0.045^{+0.164}_{-0.045}$ and $0.035^{+0.167}_{-0.035}$, cold, dense gas remains visible throughout their phase diagrams. Crucially, the Quenched populations continue to exhibit a cold, dense component, with median fractions of $0.015^{+0.344}_{-0.015}$ and $0.008^{+0.188}_{-0.008}$ for the \texttt{Efficient} and \texttt{Varying IMF} models, respectively. The broad upper tails further demonstrate that a subset of Quenched galaxies retains substantial cold, dense reservoirs. Such gas-rich quenching is consistent with emerging observational evidence for substantial molecular gas reservoirs in high-redshift quiescent systems \citep{2026arXiv260623649T}, as well as with local post-starburst analogues where large gas reservoirs coexist with suppressed star-formation efficiencies \citep{2022ApJ...929..154S}. These systems would appear quiescent in emission-line surveys yet retain detectable cold gas reservoirs accessible via molecular line observations.

These two physically distinct modes of quenching motivate a spatially resolved comparison of the gas reservoirs, which we present in Sec.~\ref{subsec:gas_profiles}.

\subsection{Radial structure of the gas reservoirs}
\label{subsec:gas_profiles}

\begin{figure*}
    \centering
    \includegraphics[width=\linewidth]{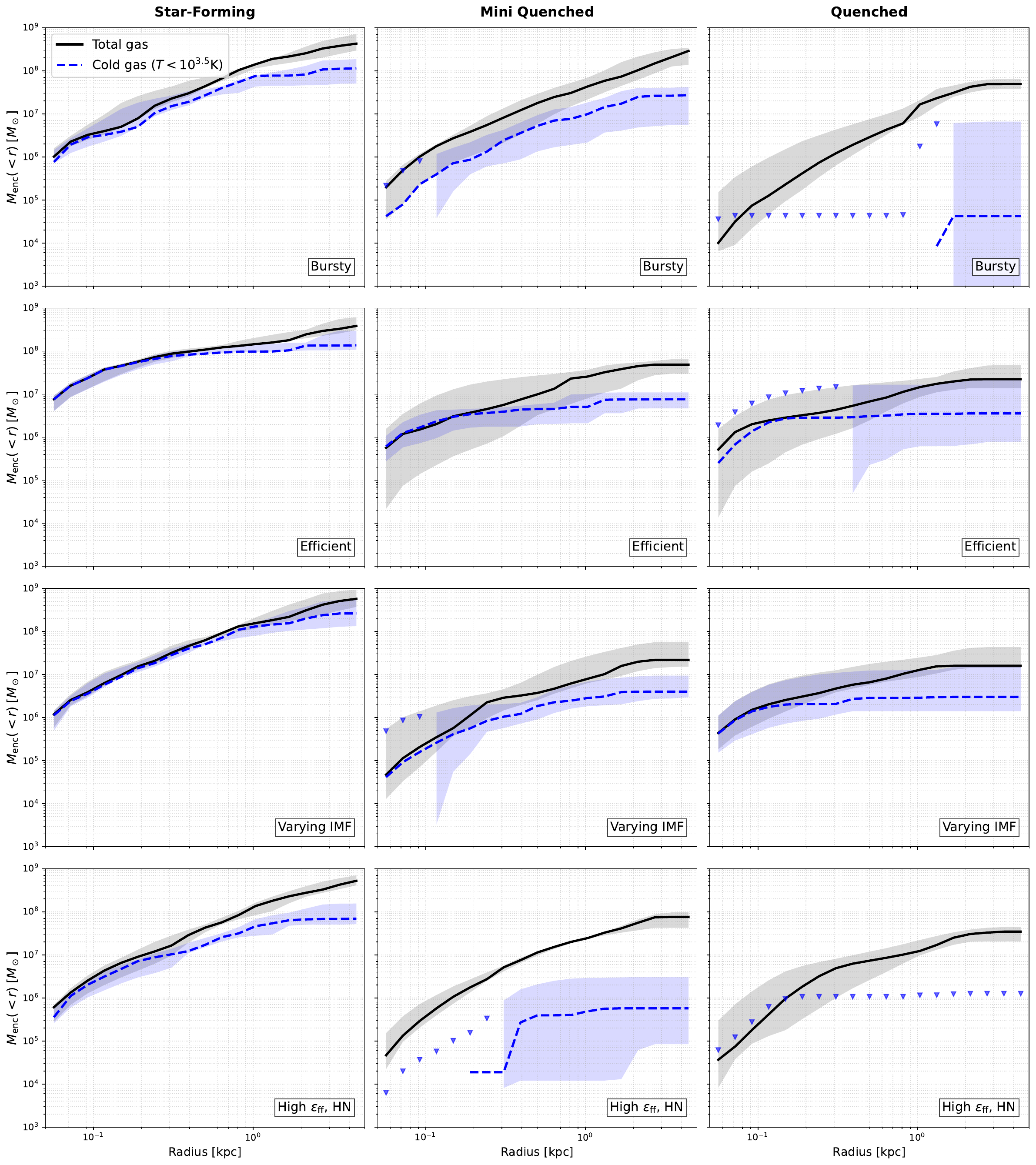}
    \caption{Enclosed gas mass profiles of galaxies in the \textsc{megatron} suite at $z\simeq8.5$. Rows correspond to the four Pop~II feedback prescriptions, while columns separate SF, MiniQ, and Quenched systems. Solid black and dashed blue curves
    show the median total and cold ($T<10^{3.5}\,\mathrm{K}$) gas masses,
    respectively, for the five systems with the largest
    $M_{200}$ in each subsample. Shaded regions show the
    1-$\sigma$ scatter. Downward triangles mark the 84th-percentile value in radial bins where the 16th percentile is zero and therefore cannot be displayed on the logarithmic axis. Where the median cold gas mass is zero, the dashed curve is absent.}
    \label{fig:mass_desnity_profile}
\end{figure*}

We next examine the radial gas profiles of the simulated galaxies
with the aim of connecting the thermodynamic differences identified in Fig.~\ref{fig:gas_phase_diagram} to the spatial distribution of gas. In Figure~\ref{fig:mass_desnity_profile} we show the cumulative enclosed total gas
mass, $M_{\mathrm{gas}}(<r)$, and cold gas mass, $M_{\mathrm{cold}}(<r)$, where cold gas is defined by $T<10^{3.5}\,\mathrm{K}$. For each feedback prescription and star-formation class, we select the five systems with the largest $M_{200}$ and calculate the median enclosed-mass profile. 

The profiles provide a spatial diagnostic of the processes associated with quiescence. A reduction in both $M_{\mathrm{gas}}(<r)$ and $M_{\mathrm{cold}}(<r)$ is consistent with feedback-driven removal or redistribution of gas from the central potential well. Recent high-redshift simulations similarly find that bursty stellar feedback can drive repeated cycles of ISM ejection and re-accretion in low-mass galaxies, temporarily depleting the central gas reservoir and suppressing star formation \citep{2025ApJ...985..126G, 2025MNRAS.544..513M}. The strength of this response is sensitive to the adopted star-formation and feedback model, with more clustered star formation capable of producing stronger outflows and more rapid disruption of dense star-forming structures \citep{2025A&A...693A.149K,2026OJAp....958199C}.

Conversely, the survival of a centrally concentrated cold component indicates that star formation can remain strongly suppressed without complete removal of the cold gas reservoir. In this case, regulation may instead reflect changes in the thermodynamic or turbulent state of the ISM, or in the efficiency with which cold gas reaches star-forming conditions, rather than simple gas depletion. This interpretation is consistent with recent high-redshift simulations showing that the density--temperature structure and star-forming state of the ISM can respond strongly to the adopted feedback and star-formation prescription even when substantial gas remains present \citep{2025MNRAS.544..513M,2026OJAp....958199C}.

The SF populations possess the largest gas reservoirs in all four models. Their median total gas masses within $4\,\mathrm{kpc}$ range from $3.84\times10^{8}$ to $5.72\times10^{8}\,\mathrm{M_\odot}$, while their cold gas masses range from $6.91\times10^{7}$ to $2.62\times10^{8}\,\mathrm{M_\odot}$. The cold gas is particularly important in the central regions: the ratios of the median cold and total enclosed masses within $0.5\,\mathrm{kpc}$ range from approximately $40$ per cent in the \texttt{High~$\epsilon_{\mathrm{ff}}$, HN} model to $81$ per cent in the \texttt{Efficient} and \texttt{Varying IMF} models, with the \texttt{Bursty} model intermediate at $62$ per cent. These centrally concentrated cold reservoirs provide the fuel required to sustain ongoing star formation.

The \texttt{Bursty} model displays a progressive, two-stage depletion of the cold component with increasing severity of quenching. Within $0.5\,\mathrm{kpc}$, the median total gas mass decreases from $4.38\times10^{7}\,\mathrm{M_\odot}$ in SF galaxies to $1.78\times10^{7}\,\mathrm{M_\odot}$ in MiniQ systems and $2.88\times10^{6}\,\mathrm{M_\odot}$ in Quenched systems. The corresponding cold gas mass falls from $2.69\times10^{7}$ to $5.24\times10^{6}\,\mathrm{M_\odot}$ and then to a median of zero, corresponding to cold fractions of $62$, $30$ and $0$ per cent. The MiniQ stage is distinguished by the gas on larger scales: within $4\,\mathrm{kpc}$ these systems still retain $2.89\times10^{8}\,\mathrm{M_\odot}$ of total gas ($68$ per cent of the SF value), even though their cold gas has fallen by a factor of approximately $4$, from $1.13\times10^{8}$ to $2.72\times10^{7}\,\mathrm{M_\odot}$. Substantial removal of gas only occurs by the Quenched stage, where the total gas mass within $4\,\mathrm{kpc}$ is a factor of approximately $9$ below that of the SF sample, the median cold gas mass is zero within $1\,\mathrm{kpc}$, and it reaches only $4.24\times10^{4}\,\mathrm{M_\odot}$ within $4\,\mathrm{kpc}$. The steep cumulative total-gas profile between $0.5$ and $2\,\mathrm{kpc}$ indicates that the little remaining gas is preferentially located at larger radii. These profiles are consistent with the amplified supernova energy first heating and displacing gas from the central region, with removal becoming substantial only once star formation has been suppressed over long timescales. This mirrors the mechanically driven quenching found in cosmological zoom simulations of dwarf galaxies subject to strong stellar feedback \citep{2019MNRAS.490.4447W}. Similar behaviour is seen in recent high-redshift simulations. In \textsc{thesan-zoom}, low-mass galaxies undergo internally driven cycles in which starbursts expel material from the ISM, temporarily suppressing star formation before the gas is re-accreted \citep{2025MNRAS.544..513M}. Notably, such suppression does not require full evacuation of the halo, since ISM-scale outflows can interrupt star formation even while accretion onto the CGM continues.

A similarly gas-poor Quenched population is produced by the \texttt{High~$\epsilon_{\mathrm{ff}}$, HN} prescription, but the depletion sets in earlier in the sequence. Within $0.5\,\mathrm{kpc}$, the median total gas mass falls from $4.28\times10^{7}\,\mathrm{M_\odot}$ in SF galaxies to $1.13\times10^{7}\,\mathrm{M_\odot}$ in MiniQ systems and $7.31\times10^{6}\,\mathrm{M_\odot}$ in Quenched systems, while the cold gas mass drops by a factor of approximately $40$ already in the MiniQ stage, from $1.71\times10^{7}$ to $3.94\times10^{5}\,\mathrm{M_\odot}$ (cold fractions of $40$ and $3.5$ per cent), and is zero in the Quenched sample. Within $4\,\mathrm{kpc}$ the total gas mass is already a factor of approximately $7$ below the SF value in the MiniQ population and a factor of approximately $15$ below it in the Quenched population ($3.47\times10^{7}$ against $5.24\times10^{8}\,\mathrm{M_\odot}$), with a median cold gas mass of zero at all radii. Unlike in the \texttt{Bursty} model, the Quenched total-gas profile is not steepened, indicating that gas has been lost from the whole $4\,\mathrm{kpc}$ region rather than only displaced outwards within it. In contrast to the \texttt{Bursty} model, where MiniQ galaxies retain most of their total gas, these systems have therefore already lost most of both their total and cold gas by the time star formation is suppressed on $10\,\mathrm{Myr}$ timescales. These profiles are consistent with an ejective mode of regulation, in which highly clustered, hypernova-assisted star formation removes or redistributes gas away from the central star-forming region. High-redshift dwarf simulations also show that more clustered star formation can generate stronger outflows and disrupt dense star-forming structures more rapidly \citep{2025A&A...693A.149K}. The depleted central reservoirs in the \texttt{Bursty} and \texttt{High~$\epsilon_{\mathrm{ff}}$, HN} models are qualitatively consistent with this feedback-regulated, ejective picture.

The \texttt{Efficient} and \texttt{Varying IMF} models produce qualitatively different profiles. Quenched galaxies in these models retain median cold gas masses of $3.10\times10^{6}$ and $2.86\times10^{6}\,\mathrm{M_\odot}$ within $0.5\,\mathrm{kpc}$, respectively. These values correspond to approximately $46$ and $43$ per cent of their median central gas masses. Moreover, the cold gas profiles are nearly flat outside the central region: approximately $87$ per cent and $95$ per cent of the cold mass enclosed within $4\,\mathrm{kpc}$ is already contained within $0.5\,\mathrm{kpc}$ in the \texttt{Efficient} and \texttt{Varying IMF} models, respectively. Quiescence in these systems therefore does not require complete removal of the temperature-selected cold reservoir. When considered alongside the phase-space distributions in Fig.~\ref{fig:gas_phase_diagram}, these profiles instead favour suppression of star formation within a retained, compact cold component, consistent with local post-starburst analogues in which large gas reservoirs coexist with suppressed star-formation efficiencies \citep{2022ApJ...929..154S}, and with emerging observational evidence for substantial molecular gas reservoirs in high-redshift quiescent systems \citep{2026arXiv260623649T}.

The radial profiles therefore separate the Quenched population into two broad structural regimes. The \texttt{Bursty} and \texttt{High~$\epsilon_{\mathrm{ff}}$, HN} populations are cold-gas poor and show evidence for strong central depletion, whereas Quenched galaxies in the \texttt{Efficient} and \texttt{Varying IMF} models can retain compact cold reservoirs. This distinction provides a spatial counterpart to the model-dependent duty-cycle behaviour discussed in Sec.~\ref{subsec:duty_cycle}. In particular, the strong stellar-mass dependence of $f_{\rm duty}$ for the Quenched populations in the \texttt{Efficient} and \texttt{Varying IMF} runs indicates that progressively more massive systems remain actively SF for a larger fraction of their available lifetime before entering the Quenched state. The survival of centrally concentrated cold gas in these same models suggests that this increased duty cycle is enabled by the continued availability of a gaseous reservoir, rather than simply by resistance to complete baryonic evacuation.

Taken together, the duty cycles and radial gas profiles therefore point to a close connection between the temporal and structural modes of feedback regulation. In models where sustained quiescence is accompanied by strong depletion of the cold reservoir, feedback can terminate star formation through direct heating, removal, or redistribution of the available fuel. By contrast, where compact cold gas survives, sustained quiescence can arise despite the continued presence of nominally star-forming material, with stellar mass controlling how long that reservoir sustains star formation before suppression becomes effective. Feedback physics consequently determines not only how frequently and for how long galaxies are active, but also whether their eventual quiescence is associated with removal of the cold gas supply or suppression of star formation within retained gas.

\subsection{Observational Signatures}
\label{subsec:observational_signatures}

A key goal of this work is to connect the physical quenching mechanisms identified above with observational diagnostics accessible to current and future high-redshift surveys. More broadly, this provides a means of distinguishing between the different physical solutions proposed within the \textsc{megatron} framework to reproduce the observed high-redshift galaxy population. These models invoke different prescriptions for star formation and stellar feedback, which imprint themselves on the density and thermodynamic structure of the ISM, as demonstrated in Fig.~\ref{fig:gas_phase_diagram}, and consequently on the emergent observable properties of high-redshift galaxies \citep{2025arXiv251005201K, 2026OJAp....956097K, 2026OJAp....958199C}; see also \textcolor{blue}{Choustikov et al., Matsumoto et al., and Quilt et al. (in prep.)}. In this section, we therefore examine how these differences propagate into three complementary observational signatures of the different quenching states: the Balmer break strength, the UV continuum slope $\beta_{\rm UV}$, and the full spectral energy distributions (SEDs).

\subsubsection{The Balmer Break--UV Slope Plane in \textsc{megatron}}
\label{subsubsec:balmer_break_physics}

\begin{figure}
    \includegraphics[width=\columnwidth]{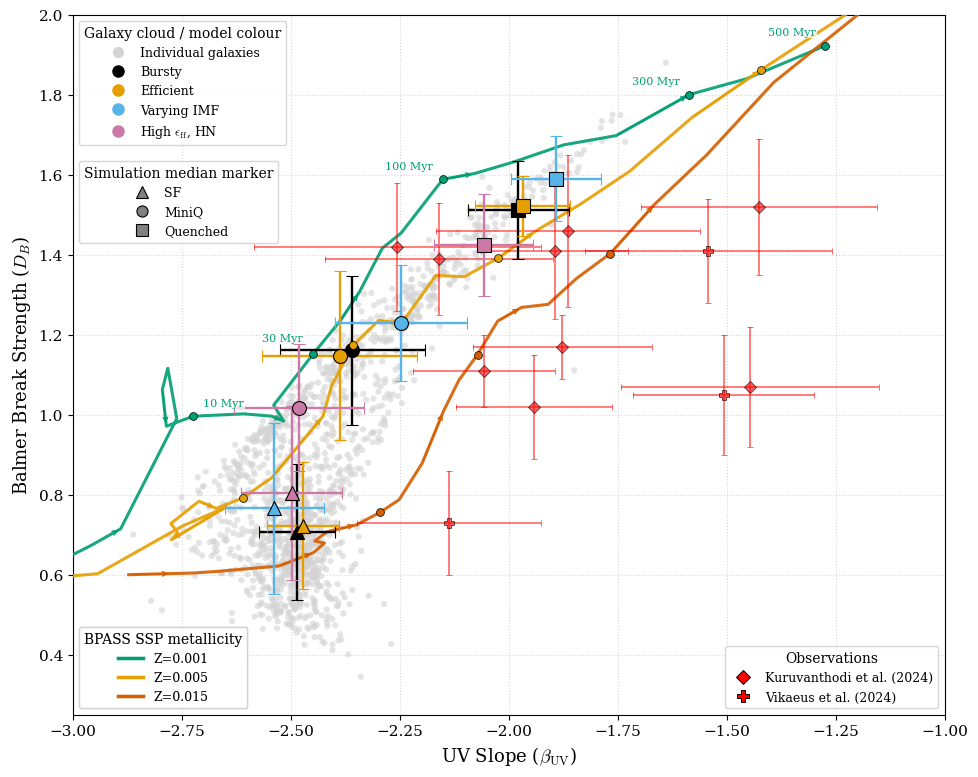}
    \caption{Balmer break strength, $D_B$, as a function of the rest-frame UV spectral slope, $\beta_{\rm UV}$, for galaxies in the four \textsc{megatron} simulations at $z=8.5$. Light-grey points show individual simulated galaxies. Coloured symbols show the median relation for each feedback prescription, with colours denoting the \texttt{Bursty} (black), \texttt{Efficient} (orange), \texttt{Varying IMF} (blue), and \texttt{High $\epsilon_{\mathrm{ff}}$, HN} (pink) models. Marker shapes identify the galaxy population: triangles, circles, and squares correspond to actively SF, MiniQ, and Quenched galaxies, respectively. Horizontal and vertical error bars indicate the $1\sigma$ scatter in $\beta_{\rm UV}$ and $D_B$ within each model and population. Solid curves show the passive evolution of BPASS v2.2.1 binary SSPs with $Z=0.001$, $0.005$, and $0.015$, evaluated using the same spectral-index definitions; labelled ages are shown along the $Z=0.001$ track. Red symbols show observational measurements from \citet{2024A&A...691A.310K} and \citet{2024MNRAS.529.1299V}, represented by diamonds and pentagons, respectively.}

    \label{fig:balmer_break}
\end{figure}

The Balmer break and UV slope are well-established probes of stellar population age and star-formation history, and are directly measurable with JWST spectroscopy at $z\gtrsim8$. The Balmer break strength, $D_B$, traces the change in the stellar continuum across the Balmer series limit at $3646$~\AA\ and is sensitive to the contribution of intermediate-age stellar populations. The rest-frame UV slope, $\beta_{\rm UV}$, instead traces the shape of the ultraviolet continuum dominated by young massive stars, with more negative values corresponding to bluer spectra. It is therefore sensitive to recent star formation, while also being affected by dust attenuation and nebular continuum emission. We define the Balmer-break
strength as 

\begin{equation}
D_B =
\frac{\left\langle f_\lambda\right\rangle_{4150-4250\,\text{\AA}}}
     {\left\langle f_\lambda\right\rangle_{3400-3600\,\text{\AA}}},
\end{equation}
following \citet{2019MNRAS.489.3827B}. We measure the rest-frame UV continuum slope by fitting a power law, $f_\lambda\propto\lambda^{\beta_{\rm UV}}$, over $1340\leq\lambda\leq2700$~\AA. We exclude the wavelength intervals $1440$--$1590$~\AA, $1620$--$1680$~\AA, and $1860$--$1980$~\AA\ to avoid emission-line contamination, following the methodology of \citet{2026MNRAS.548ag808S}. The \textsc{megatron} spectra used for both measurements include the nebular continuum, such that the derived $D_B$ and $\beta_{\rm UV}$ incorporate both the underlying stellar continuum and its reprocessing by the surrounding ionised gas.

In Fig.~\ref{fig:balmer_break} we show the Balmer break strength against $\beta_{\rm UV}$ for all individual galaxies across the four \textsc{megatron} models, alongside the combined population medians and their associated $1\sigma$ scatter. We additionally show the passive evolution of BPASS v2.2.1 binary simple stellar populations (SSPs) at $Z=0.001$, $0.005$, and $0.015$ \citep{2017PASA...34...58E,2018MNRAS.479...75S}. The SSPs are analysed using exactly the same definitions of $\beta_{\rm UV}$ and $D_B$ as the simulated galaxies, with labelled points indicating age along the $Z=0.001$ track. These tracks provide a reference for the direction and approximate magnitude of spectral evolution expected from passive stellar ageing and illustrate its metallicity dependence. They should not be interpreted as direct evolutionary tracks for the \textsc{megatron} galaxies.

For comparison, we also include observational data points from \citet{2024A&A...691A.310K} and \citet{2024MNRAS.529.1299V}. We extracted their raw spectra from the DAWN JWST Archive \citep{brammer_2025_15472354} and derived the $\beta_{\rm UV}$ slopes using the same method as for our simulated sample. However, caution is advised when comparing these directly, as the simulated galaxies probe lower stellar masses and their mock spectra are not dust-corrected.

The three galaxy populations populate this diagnostic plane in a physically intuitive manner. Star-forming galaxies, with their young, hot stellar populations, cluster in the lower-left region of the diagram (weak Balmer break, blue UV slope). Quenched galaxies, whose optical light is dominated by older stellar populations as expected when little star formation has occurred recently, occupy the upper-right region (strong Balmer break, red UV slope). MiniQ galaxies fill the intermediate region of the diagram, reflecting the diversity of their assembly and quenching histories. Importantly, this SF--MiniQ--Quenched sequence is broadly aligned with the direction of the BPASS SSP ageing track.

Across all four models, SF galaxies occupy a narrow locus of blue UV slopes and weak Balmer breaks. Small offsets may reveal the imprint of both the adopted stellar populations and the subgrid physics. The \texttt{Varying IMF} model produces the bluest median UV slope among SF galaxies ($\beta_{\rm UV}=-2.523$). The origin of this offset is not attributable to ISM clearing alone. Unlike the other prescriptions, the \texttt{Varying IMF} model employs stellar SEDs that vary consistently with its density- and metallicity-dependent IMF, directly modifying the intrinsic UV spectrum and ionizing photon production. The latter also alters the contribution of nebular continuum emission, which can substantially redden the emergent UV spectrum when ionizing photons are efficiently reprocessed \citep{2025OJAp....8E.104K}. Consequently, the observed $\beta_{\rm UV}$ reflects a combination of the intrinsic stellar SED and the subsequent processing of that radiation by the surrounding ISM.

However, feedback-driven disruption of star-forming regions remains important. The enhanced ionizing and feedback output associated with the varying IMF can reduce the density of the surrounding gas and increase the effective leakage of ionizing photons, thereby suppressing nebular reprocessing and allowing a bluer stellar continuum to emerge \citep{2026arXiv260936124K}. However, the \texttt{High $\epsilon_{\mathrm{ff}}$, HN} model clears the ISM even more rapidly than the \texttt{Varying IMF} model \textcolor{blue}{(Choustikov et al. in prep.)}, demonstrating that the bluest slopes of the latter cannot be interpreted simply as a consequence of more efficient gas clearing. Instead, they arise from the coupled effects of its distinct IMF-dependent stellar SEDs, enhanced ionizing photon production, and feedback-regulated nebular reprocessing.

In contrast to the UV slope, the Balmer break strengths of SF galaxies show only a weak ordering across models. The \texttt{Bursty} model yields the weakest median $D_B=0.692$ among the SF population, closely followed by the \texttt{Efficient} model ($D_B=0.708$), with the \texttt{Varying IMF} ($D_B=0.751$) and \texttt{High $\epsilon_{\mathrm{ff}}$, HN} ($D_B=0.790$) models slightly higher. These offsets are small compared with the $1\sigma$ scatter within each population and should be regarded as tentative.

The uniformly weak Balmer breaks of the SF populations ($D_B\approx0.7$--$0.8$ in all models) are primarily consistent with their very young stellar populations. The youngest stars contribute strongly to the continuum blueward of the Balmer edge, while the intermediate-age stellar population responsible for a pronounced Balmer break has not yet come to dominate the integrated spectrum. This interpretation is qualitatively consistent with \citet{2024ApJ...967..172S}, who analyse \textsc{sphinx} star-formation histories at $z=8$--10 and find that most galaxies exhibit continuous or rising star formation, while extended feedback-induced interruptions are rare. Such prolonged periods of suppressed star formation allow the youngest stellar populations to fade and stronger Balmer breaks to develop.

Nebular reprocessing provides an additional complication. The nebular continuum itself can weaken the Balmer break by enhancing the continuum blueward of the Balmer limit, particularly in very young galaxies \citep{2024MNRAS.527.7965W}, and the degree of reprocessing depends on how efficiently the surrounding gas is cleared and on the escape of ionizing photons \citep{2026OJAp....958199C}. The observed $D_B$ therefore reflects the competing effects of stellar-population age, star-formation history, and feedback-regulated nebular emission, rather than being a direct measure of ISM clearing alone.

Looking at the MiniQ population, the most striking result is the essentially null evolution of the UV slope in the \texttt{High $\epsilon_{\mathrm{ff}}$, HN} model. SF and MiniQ galaxies have median UV slopes of $\beta_{\rm UV} = -2.483$ and $-2.480$, respectively, a difference of $\Delta\beta = 0.003$ that is negligible compared to the scatter within each population. The corresponding Balmer break step is also the smallest ($\Delta D_B = 0.23$, from $0.790$ to $1.018$). This is consistent with the quenching mechanism in this model, in which $\epsilon_{\rm ff}=100\%$ and a minimum stellar particle mass of $2000\,\mathrm{M}_\odot$ concentrate star formation into short, spatially clustered, hypernova-assisted events that clear the central ISM abruptly (Sec.~\ref{subsec:gas_profiles}; \citealp{2025A&A...693A.149K}). If star formation is suppressed on a timescale shorter than the main-sequence lifetime of the stars formed during the preceding burst, the OB population that powered the blue UV slope is still alive and luminous when the galaxy is classified as MiniQ, and the UV continuum retains the imprint of the burst. This points to a potential degeneracy for JWST spectroscopic surveys; a galaxy in this regime may appear, from its UV colour alone, to be actively forming stars when it is, in fact, in a phase of suppressed activity.

At the opposite extreme in $\beta_{\rm UV}$, the \texttt{Varying IMF} model shows the largest SF-to-MiniQ shift in UV slope, $\Delta\beta_{\rm UV}=+0.276$ (from $-2.523$ to $-2.247$), accompanied by a Balmer break step of $\Delta D_B=+0.479$ (from $0.751$ to $1.230$) that is comparable to those of the \texttt{Efficient} and \texttt{Bursty} models. Part of this shift is naturally associated with the rapid fading of the youngest, UV-bright stellar population once star formation is suppressed. However, the evolution in $\beta_{\rm UV}$ should not be interpreted as a purely stellar-evolution signature of a top-heavy burst. In this prescription, the top-heavy IMF increases the ionizing photon budget, making the observed UV slope especially sensitive to how efficiently those photons are reprocessed by the surrounding gas. As the local ISM evolves following the burst, changes in gas density, escape fraction, and non-equilibrium nebular emission can amplify or suppress the apparent reddening of the UV continuum. The large displacement of the \texttt{Varying IMF} MiniQ population in the $D_B$--$\beta_{\rm UV}$ plane therefore reflects a combination of recent star-formation suppression, fading of the youngest stellar component, and ISM-regulated nebular reprocessing.

\begin{figure*}
    \centering
    \includegraphics[width=\linewidth]{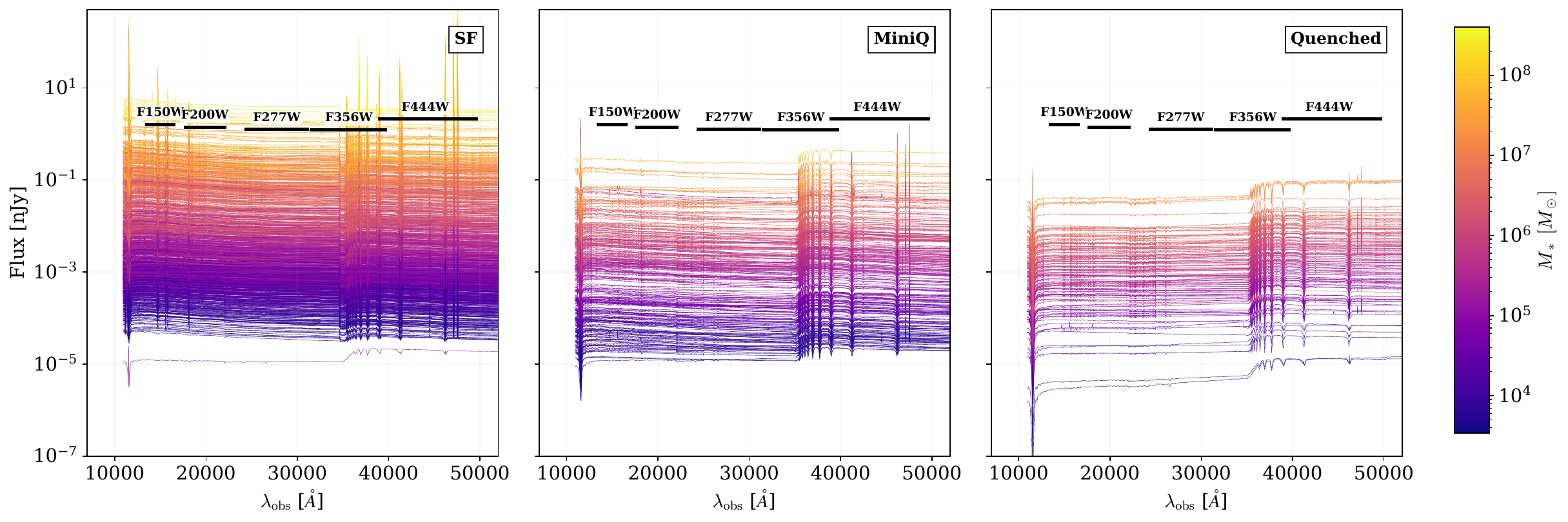}
    \caption{
    Observed-frame SEDs of all galaxies in the \textsc{megatron} simulations at $z=8.5$, grouped by star-formation state: SF (left), MiniQ (centre), and Quenched (right). Each curve represents the integrated, dust-free spectrum of an individual galaxy, including both the stellar continuum and the nebular continuum, converted to observed flux density and shifted to the observer frame. Spectra are colour-coded by total stellar mass, as indicated by the colour bar. Horizontal black bars denote the wavelength coverage and approximate point-source sensitivity limits of \textit{JWST}/NIRCam broad-band filters for a $20\,\mathrm{hr}$ exposure
    at $\mathrm{S/N} \simeq 3$.}
    \label{fig:SEDs}    
\end{figure*}

The \texttt{Efficient} and \texttt{Bursty} models occupy an intermediate regime, with SF to MiniQ steps of $\Delta\beta = +0.069$ and $+0.111$, and $\Delta D_B = +0.439$ and $+0.469$, respectively. Despite the different feedback physics, weak fiducial supernovae in the \texttt{Efficient} model versus $5\times$ amplified supernovae in the \texttt{Bursty} model, the similar evolutionary steps suggest that the spectroscopic and photometric signatures of quenching onset are governed more by the typical age of the MiniQ stellar population than by the mechanism of quenching itself, provided the quenching timescale is longer than the OB lifetime (i.e.\ not as impulsive as in the \texttt{High $\epsilon_{\mathrm{ff}}$, HN} case). In both models, the UV slope of MiniQ galaxies is already modestly redder than the SF population, indicating that a meaningful fraction of the short-lived UV-bright stars have evolved away by the time quenching is detected.

Finally, across all four prescriptions, Quenched galaxies converge toward a common spectral character: strong Balmer breaks ($D_B \approx 1.44$--$1.60$) and modestly red UV slopes ($\beta_{\rm UV} \approx -1.91$ to $-2.07$). This convergence primarily reflects the ageing of the stellar populations once star formation is quenched, most clearly traced by the growth of the Balmer break. The UV slope, however, remains sensitive to residual nebular continuum emission and to the density structure of the surrounding gas, and therefore should not be interpreted as a pure stellar-age indicator.

The \texttt{Varying IMF} model produces the strongest median Balmer break in the Quenched sample ($D_B = 1.604$) and the reddest UV slope ($\beta_{\rm UV} = -1.906$). The strong break reflects a particularly clear stellar-population ageing
signature following quenching. Once the massive-star-dominated burst population has fully evolved away, the remaining intermediate-mass stellar population produces a pronounced hydrogen Balmer continuum step. The total evolutionary range in the Balmer break from SF to Quenched is $\Delta D_B = 0.853$ in the \texttt{Varying IMF} model, the largest of any prescription, and the full $\beta_{\rm UV}$ range of $\Delta\beta = +0.617$ likewise exceeds the other models. This large dynamic range makes the \texttt{Varying IMF} model the most optimistic scenario for photometric selection of quenched galaxies: the contrast between SF and Quenched galaxies is maximised relative to the other prescriptions.

The \texttt{High $\epsilon_{\mathrm{ff}}$, HN} model produces the weakest Quenched Balmer break ($D_B = 1.439$) and the most negative UV slope ($\beta_{\rm UV} = -2.072$) among the Quenched populations, and its Quenched galaxies have a smaller total evolutionary range ($\Delta D_B = 0.649$, $\Delta\beta = +0.411$) than in the other models. This reduced range arises almost entirely in the SF-to-MiniQ step discussed above: the MiniQ-to-Quenched step in this model ($\Delta D_B = 0.421$, $\Delta\beta = +0.408$) is comparable to that of the other prescriptions. The differences between the Quenched populations are modest compared with the $1\sigma$ scatter, and we do not interpret them further.

In contrast to the model-dependent SF-to-MiniQ shift in $\beta_{\rm UV}$, the offset between the MiniQ and Quenched populations is similar across all four prescriptions. This convergence signals that once a galaxy has entered the MiniQ phase and its younger stellar population has faded, the subsequent evolution toward the Quenched regime is driven by the same stellar ageing physics. The subgrid model determines how quickly and how decisively a galaxy enters the MiniQ phase, and with what initial spectral properties, but it has comparatively little influence on the subsequent spectroscopic and photometric evolution toward the fully quenched state. This result suggests that the MiniQ to Quenched transition may be a near-universal evolutionary track in the Balmer break--UV slope plane and that the diversity introduced by different feedback prescriptions is concentrated at earlier stages of the quenching sequence.

\subsubsection{SEDs and JWST Detectability}
\label{subsubsec:seds}

The simulated SEDs shown in Fig.~\ref{fig:SEDs} reveal a stark contrast in the observational accessibility of active and quiescent populations at $z \sim 8.5$. Consistent with previous theoretical work \citep{Gelli_2023}, we find that Quenched galaxies are significantly fainter than SF galaxies of equivalent stellar mass. The absence of young, massive OB stars leads to a rapid decline in rest-frame UV and optical continuum flux, placing the predicted fluxes of Quenched systems well below the sensitivity limits of \textit{JWST}/NIRCam broad-band filters across the wavelength range covered by F150W through F444W. 

While SF galaxies at the upper end of our stellar mass range remain detectable at current survey depths, the Quenched population probed in \textsc{megatron} lies systematically outside the detection horizon of all JWST programmes, and MiniQ galaxies occupy an intermediate, largely sub-threshold regime. This implies that high-redshift photometric and spectroscopic surveys based primarily on flux-selected samples may be strongly biased toward actively SF galaxies, potentially missing a substantial population of faint, quiescent low-mass systems and thereby biasing the inferred census of stellar mass assembled during the EoR. This motivates recent efforts to construct photometrically selected, approximately stellar-mass-complete samples at high redshift \citep[e.g.,][]{2025MNRAS.544.4551S}, designed to recover a more representative census of the underlying galaxy population.

Our SED modelling further highlights the absence of prominent nebular emission lines in the Quenched population. This lack of emission lines is a natural consequence of the deficit of ionising photons in the absence of young, massive stars, and constitutes an additional observational signature distinguishing Quenched galaxies from their active counterparts in slitless spectroscopic surveys.

Gravitational lensing offers a route to these otherwise undetectable systems. The GLIMPSE survey reaches observed depths of $\approx30.6$--$30.9$\,mag ($5\sigma$; $\approx1.6$--$2.1$\,nJy) behind Abell~S1063, comparable to the 20\,hr blank-field limits adopted in Fig.~\ref{fig:SEDs}, but with magnifications of $\mu>10$ for many background sources and up to $\mu\sim10^{3}$ near the critical curves \citep{2025arXiv251107542A}. A magnification of $\mu\sim10$--30 brightens a source by $\approx2.5$--3.7\,mag, bringing the brightest Quenched and MiniQ galaxies in our sample to fluxes of order 1\,nJy and therefore within reach of such programmes.

\section{Discussion}
\label{sec:discussion}

\subsection{Implications for high-redshift star-formation regulation}
\label{subsec:discussion_implications}

Recent theoretical and empirical work increasingly supports a picture in which low-mass galaxies at high redshift undergo strongly time-variable star formation, including short-lived excursions into suppressed states \citep{2025ApJ...985..126G,2025MNRAS.544..513M, 2026MNRAS.547ag415M,2026arXiv260922431S}. The outstanding question is therefore not simply whether bursty star-formation cycles can occur, but how robust their predicted incidence, duration, and physical character are to the uncertain modelling of star formation and stellar feedback.

This is the uncertainty isolated by \textsc{megatron}. Because the four simulations share identical initial conditions, numerical resolution, and large-scale environment, differences between them can be attributed directly to the adopted Pop~II star-formation and feedback prescriptions. We find that highly time-variable star formation occurs across the suite, but its manifestation is strongly model dependent. The prescriptions predict substantially different frequencies of suppressed phases, star-forming duty cycles, and physical states of the gas while the recent SFR is low. A generic prediction of burstiness therefore does not imply a robust prediction for either the abundance or the nature of temporarily suppressed galaxies.

This distinction is particularly relevant to current attempts to explain the high star-formation efficiencies and UV luminosities inferred for some early JWST galaxies. Enhanced star-formation efficiencies, stochastic star-formation histories, and IMF variations have all been explored as possible ingredients in models of the bright high-redshift population \citep{2025ApJ...980...10J,2024A&A...689A.244C, 2026MNRAS.545f2119S}, while the extent to which stellar feedback can regulate highly efficient star formation remains under active investigation \citep{2025OJAp....8E.140F}. Our results show that these modelling choices have consequences beyond the quantities most often used to calibrate early-galaxy models. Prescriptions capable of producing broadly plausible stellar masses and luminous galaxy populations can nevertheless predict very different duty cycles and probabilities for the same underlying galaxy population to be observed in a low-SFR state. The faint, temporarily suppressed population therefore provides information on feedback physics that is complementary to the abundance of UV-bright galaxies.

The physical state of the suppressed population provides a second level of discrimination. An sSFR-based classification does not uniquely specify how star formation has been regulated. Some Quenched populations are accompanied by strong depletion or outward redistribution of their central cold-gas reservoirs, whereas others retain substantial compact cold gas despite their low recent SFRs. These behaviours point to different balances between ejective and non-ejective regulation. We cannot establish classical preventative feedback from the present profiles alone, since doing so would require direct measurements of CGM cooling and accretion, but the retained reservoirs demonstrate that suppressed star formation does not require complete removal of the temperature-selected cold component. Consequently, gas content and recent SFR should be treated as complementary rather than interchangeable diagnostics of feedback regulation.

\subsection{Observational implications}
\label{subsec:disc_obs}

The model dependence identified above complicates the interpretation of individual high-redshift galaxies. Spectral appearance does not map uniquely onto instantaneous star-formation activity because UV continuum emission and the Balmer break respond over different timescales to the recent SFH and are additionally affected by stellar population and nebular physics. Recently suppressed galaxies can therefore retain the spectral imprint of their preceding star-forming episode, while more prolonged suppression allows the youngest stellar populations to fade and stronger Balmer breaks to emerge. Similar short-lived quiescent phases and their spectral consequences have been predicted in other high-redshift simulations \citep{2025ApJ...985..126G}. Identifying suppressed systems will therefore require combinations of recent-SFR indicators rather than reliance on a single UV colour or spectral index.

A more discriminating test of the feedback prescriptions is provided at the population level. In \textsc{megatron}, the predicted incidence of suppressed star formation diverges most strongly in the low-mass regime, making the suppressed fraction as a function of stellar mass a potentially direct constraint on how efficiently feedback interrupts star formation. The challenge is that these same systems are systematically fainter during their suppressed phases, so flux-limited samples preferentially recover galaxies while they are actively SF. Measurements of the intrinsic suppressed fraction will therefore require deep lensing surveys, approximately stellar-mass-complete selections, or explicit forward modelling of the observational selection function \citep{2025arXiv251107542A,2025MNRAS.544.4551S}.

More broadly, comparisons between high-redshift galaxy models and observations should move beyond time-averaged stellar masses and UV luminosities alone. Star-formation duty cycles, the abundance of temporarily suppressed systems, gas-reservoir structure, and time-sensitive spectral diagnostics contain additional information on how stellar feedback couples to the early ISM. In this sense, burstiness is not merely a source of scatter in high-redshift galaxy populations, but a potential observational constraint on the physics responsible for regulating their growth.

\subsection{Caveats and limitations}
\label{subsec:caveats}

Several limitations should be taken into account when interpreting the results presented in this paper. Most fundamentally, the four \textsc{megatron} prescriptions explored here do not span the full space of plausible models for star formation and stellar feedback at high redshift. Numerous alternative choices remain possible, including different star-formation criteria and efficiencies, supernova energetics and coupling, IMF variations, radiative-feedback prescriptions, and additional physical processes not varied in the present suite. The models considered here should therefore be viewed as a controlled sampling of distinct theoretical possibilities rather than an exhaustive bracketing of the uncertainty in high-redshift galaxy formation.

The value of this controlled comparison is instead to identify which predictions are strongly model dependent and which observational diagnostics may be useful for discriminating between different physical scenarios. Future simulations sampling a broader range of theoretical prescriptions will be required to establish whether the range of behaviours identified here encompasses that expected from the wider model space.

A second limitation is the small number of Quenched galaxies in some of the feedback prescriptions. In particular, the \texttt{Efficient} and \texttt{Varying IMF} models contain only 8 and 16 Quenched systems, respectively. Their Quenched fractions therefore carry relatively large counting uncertainties, while measurements of the internal properties of these populations are more susceptible to individual objects. This caveat is particularly relevant to the contrast in retained cold-gas reservoirs discussed above: the qualitative difference is present in the current samples, but its population-level prevalence requires confirmation with larger statistics. Larger simulated samples, either through increased volumes or additional zoom targets, will be required to establish the robustness of these trends.

Third, the \textsc{megatron} zoom region is centred on a single Milky Way-mass halo analogue, and therefore samples the high-redshift environment of only one protogalactic overdensity. Merger rates, tidal interactions, external radiation fields, and gas accretion may all differ in less overdense environments, which can have an effect on the gas inflows, stripping, and star formation in dwarf galaxies \citep{2016ApJ...831....1W}. Absolute suppressed fractions measured here should therefore not be assumed to represent the cosmic mean. Comparisons between the four prescriptions are less exposed to this limitation because they share identical initial conditions and environment, but establishing whether their relative behaviour persists elsewhere will require a more environmentally diverse set of simulations.

Fourth, our classification is based on sSFRs averaged over 10 and 100\,Myr at a single epoch. The results are qualitatively robust to the adopted threshold (Appendix~\ref{sec:quenching_threshold}), but the analysis does not directly determine the lifetime of individual suppressed episodes or whether they subsequently rejuvenate. Following matched galaxies across multiple snapshots will be required to measure these transitions directly. Accordingly, ``Quenched'' throughout this work denotes suppression over the preceding 100\,Myr, rather than necessarily permanent cessation of star formation.

Finally, the galaxy-integrated SEDs presented in Section~\ref{subsubsec:seds} do not include a self-consistent treatment of dust attenuation. For the low-mass, metal-poor suppressed systems that are the primary focus of this work, dust is not expected to dominate our qualitative conclusions regarding detectability. At higher stellar masses, however, attenuation may alter the UV slope and reduce the contrast between SF and suppressed systems in the $D_B$--$\beta_{\rm UV}$ plane. A full dust-radiative-transfer treatment is deferred to future work.

\section{Summary and Conclusions}
\label{sec:conclusions}

We have used the \textsc{megatron} high-redshift simulation suite to investigate how uncertain Pop~II star-formation and stellar-feedback physics regulates low-mass galaxies at $z=8.5$. The four simulations are evolved from identical initial conditions and at identical numerical resolution, allowing differences between their galaxy populations to be associated directly with the adopted sub-grid prescriptions. Our principal conclusions are:

\begin{enumerate}

    \item \textbf{Sub-grid feedback physics strongly controls the incidence of suppressed star formation.} The abundance and stellar-mass dependence of the Quenched population vary substantially between the four prescriptions. The \texttt{Bursty} and \texttt{High~$\epsilon_{\mathrm{ff}}$, HN} models produce substantially more Quenched systems than the \texttt{Efficient} and \texttt{Varying IMF} runs, while the \texttt{Bursty} prescription produces a distinct maximum in the Quenched fraction around $M_\star\sim10^{6.5}\,\mathrm{M_\odot}$. The predicted incidence of suppressed galaxies at Cosmic Dawn is therefore highly sensitive to the adopted feedback model.

    \item \textbf{The same suppressed classification can correspond to different physical gas states.}
    Quenched galaxies in the \texttt{Bursty} and \texttt{High~$\epsilon_{\mathrm{ff}}$, HN} prescriptions are characterised by strong central cold-gas depletion, consistent with predominantly ejective regulation. In contrast, suppressed galaxies in the \texttt{Efficient} and \texttt{Varying IMF} models can retain compact cold reservoirs. Low recent sSFR therefore does not uniquely identify either gas evacuation or a particular feedback mechanism.

    \item \textbf{These differences propagate into observable
    signatures and selection biases.} 
    Feedback-dependent SFHs produce different trajectories through the Balmer-break--UV-slope plane, while temporarily suppressed galaxies can remain spectrally similar to actively SF systems. At the same time, the lowest-mass suppressed systems are systematically difficult to detect in unlensed surveys. The suppressed fraction as a function of stellar mass, measured with sufficiently deep and selection-aware samples, therefore provides a particularly useful population-level test of high-redshift feedback models.

\end{enumerate}

The broader implication is that reproducing the stellar masses or UV luminosities of high-redshift galaxies is not sufficient to establish that a model captures how those galaxies grow. Within the same cosmological structures, plausible variations in unresolved star-formation and feedback physics produce different duty cycles, suppressed fractions, gas configurations, and observable states. Recovering the faint and intermittently star-forming dwarf-galaxy population will therefore provide an important complementary test of the physics regulating galaxy growth during the EoR.

\section*{Acknowledgements}
We thank William Roper for initiating the collaboration that led to this work and for many insightful informal discussions.

DA work was supported by the Science and Technology Facilities Council [grant number ST/W006839/1] through the DISCnet Centre for Doctoral Training. 

This work made extensive use of the dp016, dp265, dp373, and dp379 projects on the DiRAC ecosystem. This work was performed using the DiRAC Data Intensive service at Leicester, operated by the University of Leicester IT Services, which forms part of the STFC DiRAC HPC Facility (www.dirac.ac.uk). The equipment was funded by BEIS capital funding via STFC capital grants ST/K000373/1 and ST/R002363/1 and STFC DiRAC Operations grant ST/R001014/1. 
This work used the DiRAC@Durham facility managed by the Institute for Computational Cosmology on behalf of the STFC DiRAC HPC Facility (www.dirac.ac.uk). The equipment was funded by BEIS capital funding via STFC capital grants ST/P002293/1, ST/R002371/1 and ST/S002502/1, Durham University and STFC operations grant ST/R000832/1. 
This work was performed using resources provided by the Cambridge Service for Data Driven Discovery (CSD3) operated by the University of Cambridge Research Computing Service (www.csd3.cam.ac.uk), provided by Dell EMC and Intel using Tier-2 funding from the Engineering and Physical Sciences Research Council (capital grant EP/T022159/1), and DiRAC funding from the Science and Technology Facilities Council (www.dirac.ac.uk). DiRAC is part of the National e-Infrastructure.
This work has made use of the Infinity Cluster hosted by Institut d'Astrophysique de Paris and the Glamdring Cluster hosted by the University of Oxford. We thank Stephane Rouberol and Jonathan Patterson for smoothly administering these clusters. The authors also acknowledge financial support from Oriel College’s Research Fund.

We thank the developers and maintainers of NumPy \citep{harris2020array}, pynbody \citep{2013ascl.soft05002P}, yt \citep{2011ApJS..192....9T},  SciPy \citep{2020SciPy-NMeth} ,  matplotlib \citep{Hunter:2007}, pandas \citep{mckinney-proc-scipy-2010}, the Astrophysics Data Service, and the arXiv pre-print repository for providing open-source software and services that were used extensively in this work.
\section*{Data Availability}

The data used in this paper is available for any reasonable
requests to the corresponding author.



\bibliographystyle{mnras}
\bibliography{example} 




\appendix

\section{Robustness of quenching threshold}
\label{sec:quenching_threshold}
\begin{figure*}
    \centering
    \includegraphics[width=\linewidth]{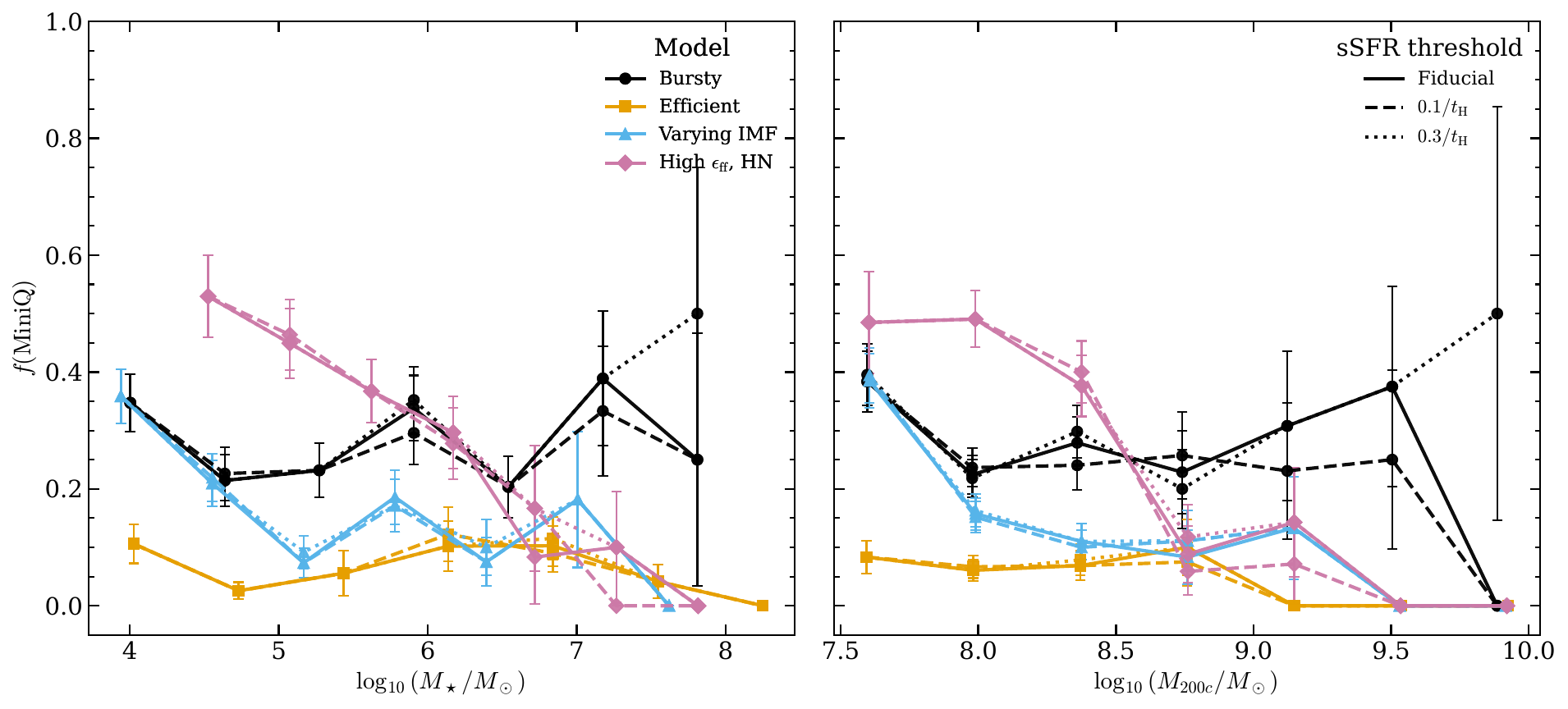}
    \caption{Sensitivity of the MiniQ galaxy fraction, $f(\mathrm{MiniQ})$, to variations in the specific star formation rate (sSFR) quenching threshold. The fraction is shown as a function of stellar mass ($\log_{10}(M_\star / \mathrm{M}_\odot)$, left panel) and halo mass ($\log_{10}(M_{\mathrm{200c}} / \mathrm{M}_\odot)$, right panel). Different models are indicated by colour and marker style as described in the legend. Line styles denote variations around our fiducial threshold, $0.2/t_{\mathrm{H}}(z)$ (solid lines), compared against a stricter $0.1/t_{\mathrm{H}}(z)$ criterion (dashed lines) and a more permissive $0.3/t_{\mathrm{H}}(z)$ criterion (dotted lines). Error bars represent binomial uncertainties. The qualitative trend across models remains structurally robust to the exact threshold choice, with minor deviations confined exclusively to low-number-count bins at the highest masses.}
    \label{fig:threshhold_MiniQ}
\end{figure*}
 
\begin{figure*}
    \centering
    \includegraphics[width=\linewidth]{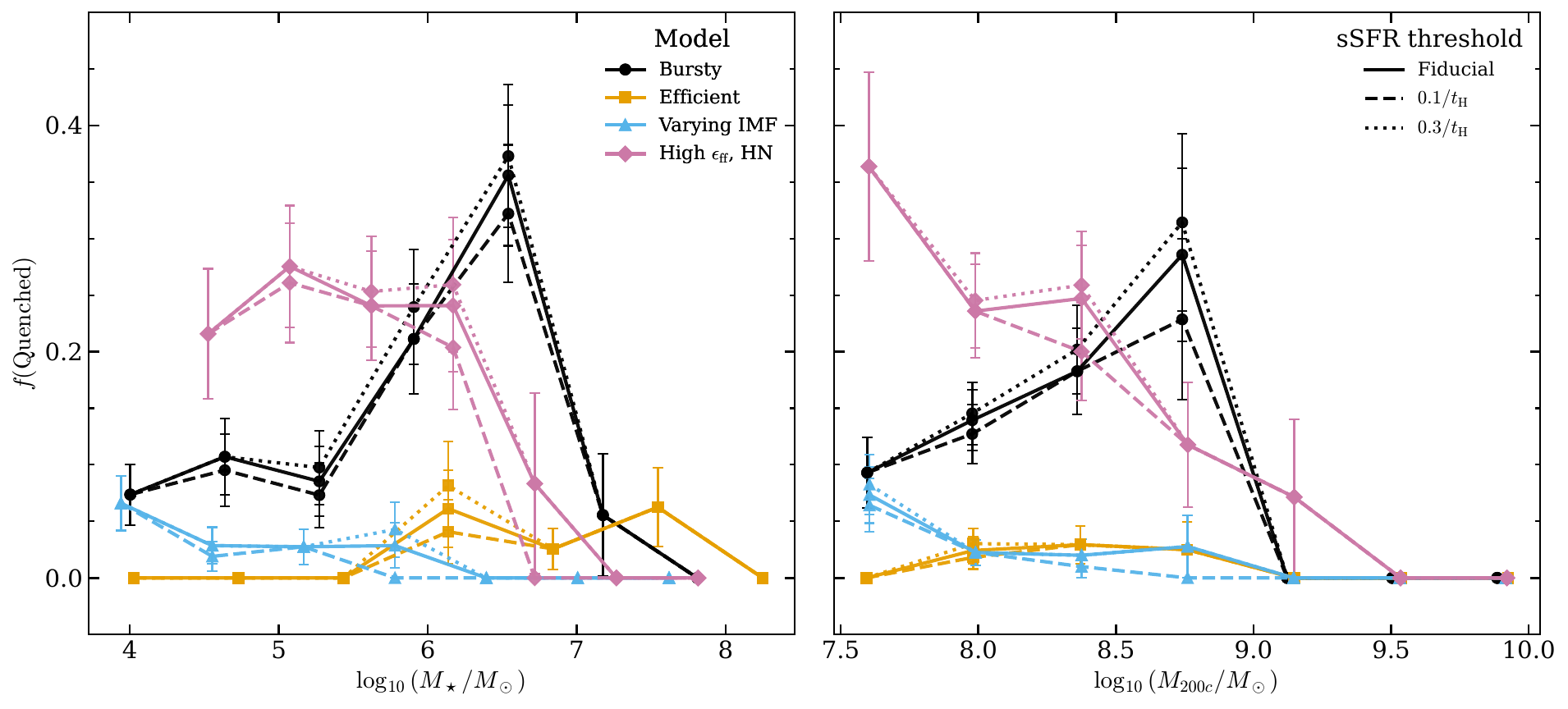}
    \caption{Same plot as Fig.~\ref{fig:threshhold_MiniQ}, but for Quenched galaxies.}
    \label{fig:threshhold_Quenched}
\end{figure*}

In this section, we present the variations in our quenched galaxy fractions under alternative classification thresholds. The results of these sensitivity tests are shown in Fig.~\ref{fig:threshhold_MiniQ} for the MiniQ population and Fig.~\ref{fig:threshhold_Quenched} for the Quenched population.

Overall, the selection of our quenched populations is highly robust against these variations. Variations in $f(\mathrm{Quenched})$ and $f(\mathrm{MiniQ})$ are minimal across the majority of the resolved mass range. Discrepancies are isolated to the highest mass bins, where the small absolute number of simulated galaxies makes them intrinsically sensitive to minor reclassifications. We therefore conclude that our choice of a fiducial $0.2 / t_H(z)$ threshold introduces no significant systematic bias to our qualitative conclusions regarding feedback efficiency and quenching timescales.

\section{Representative star-formation histories}
\label{sec:representative_sfhs}

\begin{figure*}
    \centering
    \includegraphics[width=\linewidth]{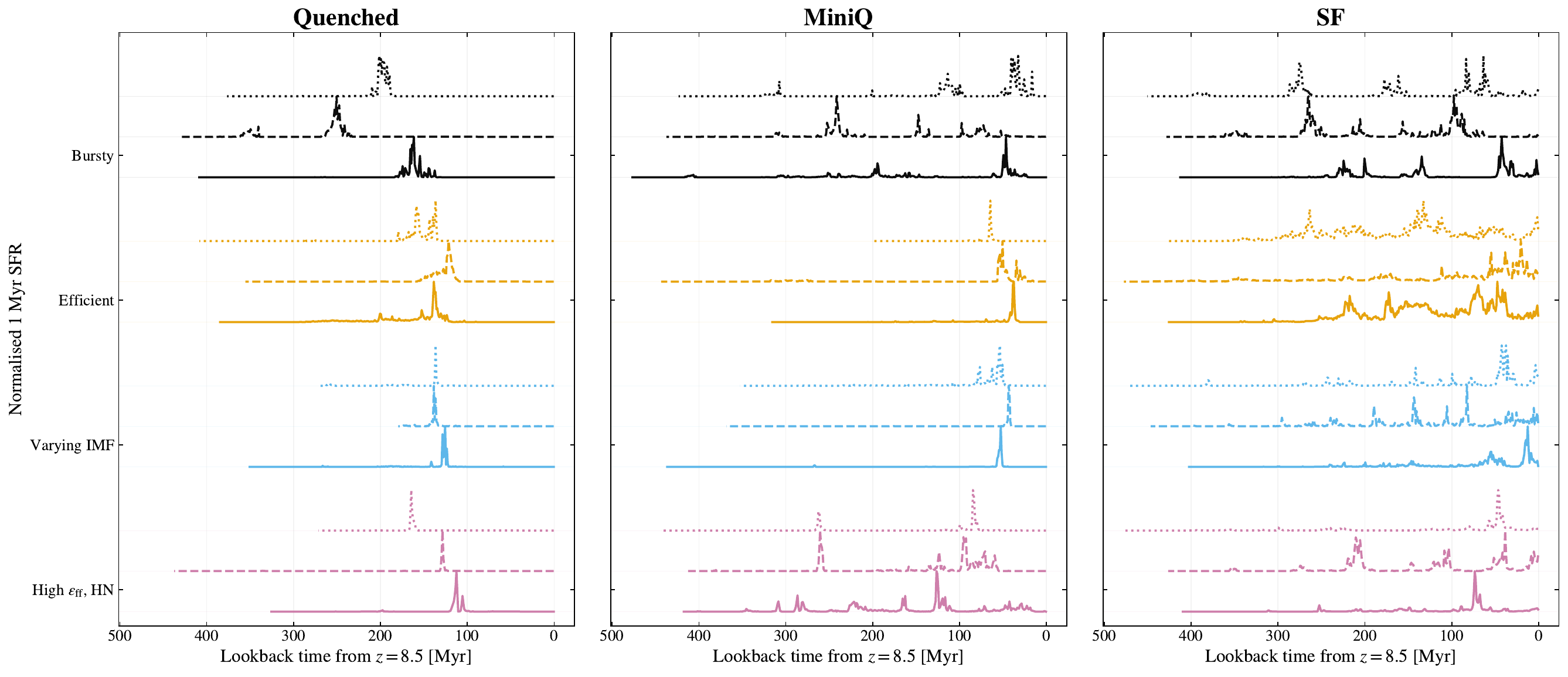}
    \caption{The 1-Myr-binned star-formation histories for galaxies in the four \textsc{megatron} feedback models at $z=8.5$. From left to right, panels show Quenched, Mini-Quenched (MiniQ), and Star-Forming (SF) galaxies. For each feedback prescription and star-formation class, we show the three galaxies with the largest final stellar masses. Solid, dashed, and dotted curves correspond to the first-, second-, and third-most massive systems, respectively. Each star-formation history is calculated in $1\,\mathrm{Myr}$ bins and independently normalised by its maximum SFR before being vertically offset for clarity.}
    \label{fig:representative_sfhs}
\end{figure*}

To illustrate the temporal behaviour underlying the duty-cycle measurements presented in Sec.~\ref{subsec:duty_cycle}, Fig.~\ref{fig:representative_sfhs} shows representative star-formation histories for the resolved \textsc{megatron} population. For each feedback prescription and star-formation class, we select the three galaxies with the largest stellar masses and reconstruct their star-formation histories in $1\,\mathrm{Myr}$ bins from the formation times and masses of their stellar particles. Each history is normalised by its own maximum SFR and vertically offset for clarity. Consequently, the figure is intended to compare the temporal structure of star formation rather than its absolute amplitude.

The $1\,\mathrm{Myr}$ binning corresponds directly to the definition of the duty cycle used in Eq.~\ref{eq:fduty}: every bin containing non-zero star formation contributes $1\,\mathrm{Myr}$ to $t_{\rm on}$, whereas intervals with zero SFR do not. The figure therefore provides a direct visual counterpart to the integrated statistic $f_{\rm duty}$. Galaxies whose star formation is concentrated into a small number of short episodes spend only a small fraction of the interval between first star formation and observation in an active state and consequently have low duty cycles. Conversely, histories populated by repeated or extended periods of star formation yield larger values of $t_{\rm on}$ and hence higher $f_{\rm duty}$.

\section{Selection effects in stellar-mass assembly times}
\label{sec:assembly_selection_test} In Sec.~\ref{subsec:mass_assembly}, we motivate a test of whether apparent differences in stellar-mass assembly between Quenched, MiniQ, and SF galaxies reflect genuinely distinct growth histories or arise partly from the recent-star-formation criteria used to define the samples. Here we present the full selection-effect test.

To quantify the magnitude of this selection effect, we construct a null comparison from the SF population. For every SF galaxy, we remove all stellar mass formed during the final 100\,Myr before $z=8.5$ and recompute both its final stellar mass and its assembly times. We define $\tau_{25}$, $\tau_{50}$, and $\tau_{75}$ as the lookback times at which 25, 50, and 75 per cent of the resulting stellar mass had been assembled, normalised by the age of the Universe at $z=8.5$. Larger values of $\tau$ therefore correspond to earlier stellar-mass assembly. From this truncated SF population we construct a comparison sample with the same stellar-mass distribution and feedback-model composition as the Quenched population. This provides a conservative null test because the Quenched criterion requires only a sufficiently low 100-Myr-averaged sSFR, whereas the comparison removes all star formation during this interval.

\begin{figure*}
    \centering
    \includegraphics[width=\linewidth]{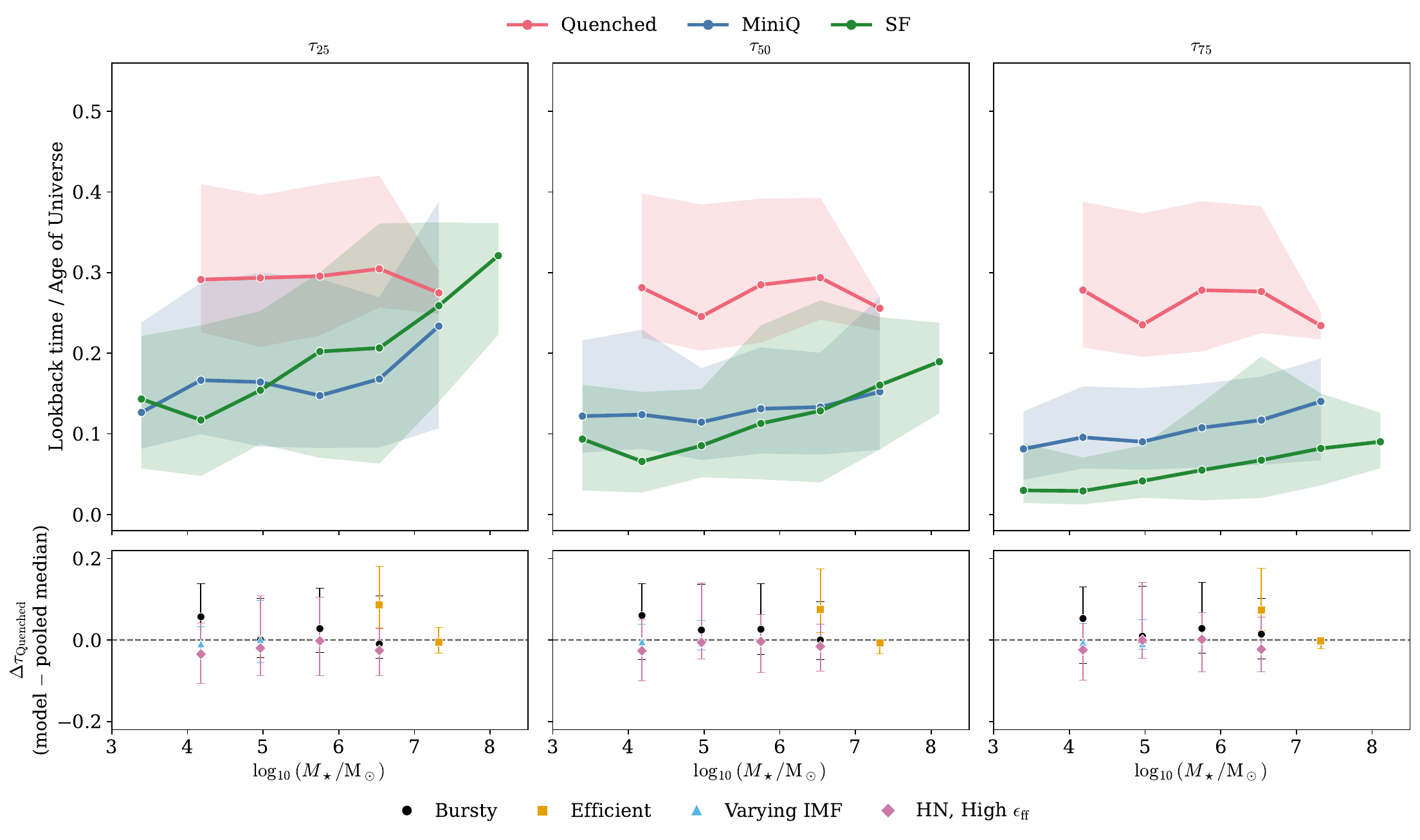}
    \caption{
    Normalised stellar assembly lookback times as a function of stellar mass for galaxies at $z=8.5$. Columns show $\tau_{25}$, $\tau_{50}$, and $\tau_{75}$, the lookback times at which a galaxy had assembled 25, 50, and 75 per cent of its $z=8.5$ stellar mass, normalised by the age of the Universe at $z=8.5$, such that larger values indicate earlier assembly.
    \textit{Top row:} median relations for Quenched (red), MiniQ (blue), and SF (green) galaxies, pooling all four \textsc{megatron} feedback prescriptions within each class; shaded regions show the $1\sigma$ in each stellar-mass bin. The dashed curve shows the median relation for SF galaxies whose star formation during the final 100\,Myr before $z=8.5$ has been removed, with assembly times and stellar masses recomputed from the truncated SFHs, and drawn with the same stellar-mass distribution and feedback-model composition as the Quenched sample.
    \textit{Bottom row:} residuals of each feedback model's Quenched median relation relative to the pooled Quenched relation; error bars show the $1\sigma$ scatter. The dashed horizontal line marks zero residual.
    }
    \label{fig:assembly_times}
\end{figure*}

The raw comparison in the top row of Fig.~\ref{fig:assembly_times} shows the expected ordering: Quenched galaxies reach $\tau_{25}$, $\tau_{50}$, and $\tau_{75}$ earlier than the SF population at fixed stellar mass, with the largest separation occurring at $\tau_{75}$. MiniQ galaxies generally occupy an intermediate locus. Taken at face value, this could be interpreted as evidence that galaxies entering the Quenched population follow a systematically earlier stellar-mass assembly pathway.

The matched null comparison demonstrates that most of this difference is instead produced by the recent-inactivity selection. After the final 100\,Myr of star formation is removed from the SF SFHs, the resulting $\tau_{50}$ and $\tau_{75}$ relations move substantially towards the Quenched population and are consistent with it within the scatter across the stellar-mass range probed here. The strong late-time assembly offset seen in the unmodified comparison is therefore largely a consequence of selecting galaxies that, by definition, have formed little or no stellar mass recently.

The behaviour of $\tau_{25}$ provides an additional check on this interpretation. Rather than retaining evidence for an earlier onset of stellar-mass growth, the truncated SF comparison generally lies at larger $\tau_{25}$ than the Quenched population. The Quenched galaxies therefore do not appear to begin assembling their stellar mass unusually early relative to the null expectation. Instead, their stellar mass is often accumulated through one or a small number of more concentrated episodes, as illustrated by the representative star-formation histories in Appendix~\ref{sec:representative_sfhs}. There is consequently no evidence from these assembly-time metrics that Quenched galaxies occupy a distinct, systematically earlier stellar-mass assembly pathway beyond that required by the absence of recent star formation.

We additionally test whether the assembly histories of Quenched galaxies depend on the adopted feedback prescription. The bottom row of Fig.~\ref{fig:assembly_times} shows the residual median $\tau_{25}$, $\tau_{50}$, and $\tau_{75}$ relations for each model relative to the pooled Quenched population. No significant systematic offset is apparent between the four prescriptions: the model-to-model differences remain small compared with the intrinsic scatter of the galaxy population. This contrasts with the strong model dependence of the star-forming duty cycles and of the time since the most recent star-formation episode discussed in Sec.~\ref{subsec:mass_assembly}.

This test highlights an important distinction between integrated assembly metrics and the temporal structure of recent star formation. Quantities such as $\tau_{50}$ and $\tau_{75}$ can differ strongly between samples selected according to their current or recent star-formation state even when there is no underlying difference in their earlier growth histories. Assembly-time comparisons between active and suppressed high-redshift galaxies should therefore be interpreted together with the timescale and threshold used to define the samples. In the present simulations, feedback leaves a much clearer imprint on the intermittency and persistence of star formation than on a distinct stellar-mass assembly trajectory into the Quenched state.

\bsp	
\label{lastpage}
\end{document}